\documentclass[a4paper,numberwithinsect,cleveref,thm-restate]{lipics-v2021}
\hideLIPIcs
\nolinenumbers

\usepackage{mathtools}
\usepackage{mathpartir}
\usepackage{placeins}
\usepackage{microtype}

\newcommand{\logicname}[1]{\ensuremath{\mathsf{#1}}}
\newcommand{\funcname}[1]{\ensuremath{\mathsf{#1}}}

\newcommand{\bN}{\mathbb{N}}
\newcommand{\bL}{\mathbb{L}}
\newcommand{\cC}{\mathcal{C}}
\newcommand{\cR}{\mathcal{R}}
\newcommand{\cF}{\mathcal{F}}
\newcommand{\cV}{\mathcal{V}}
\newcommand{\cM}{\mathcal{M}}
\newcommand{\cN}{\mathcal{N}}
\newcommand{\cP}{\mathcal{P}}
\newcommand{\cQ}{\mathcal{Q}}
\newcommand{\cT}{\mathcal{T}}
\newcommand{\cL}{\mathcal{L}}
\DeclarePairedDelimiter{\set}{\lbrace}{\rbrace}
\DeclarePairedDelimiter{\tuple}{\langle}{\rangle}
\DeclarePairedDelimiter{\size}{\lvert}{\rvert}
\DeclarePairedDelimiterX{\setpred}[2]{\lbrace}{\rbrace}{#1\,\delimsize\vert\,#2}
\newcommand{\powerset}[1]{\mathcal{P}(#1)}

\newcommand{\FO}{\logicname{FO}}
\newcommand{\SO}{\logicname{SO}}
\newcommand{\HO}{\logicname{HO}}
\newcommand{\LFP}{\logicname{LFP}}
\newcommand{\HFL}{\logicname{HFL}}
\newcommand{\PHFL}{\logicname{PHFL}}
\newcommand{\Coh}{\logicname{Coh}}
\newcommand{\HORS}{\logicname{HORS}}
\newcommand{\REG}{\logicname{REG}}
\newcommand{\EXPTIME}{\logicname{EXPTIME}}
\newcommand{\EXPSPACE}{\logicname{EXPSPACE}}
\newcommand{\DTIME}{\logicname{DTIME}}

\newcommand{\rk}{\mathrm{rk}}
\newcommand{\ord}{\mathrm{ord}}

\newcommand{\ordequiv}{\equiv_{\mathrm{ord}}}
\DeclareMathOperator{\lfpop}{lfp}
\DeclareMathOperator{\Inv}{Inv}
\DeclareMathOperator{\Gen}{Gen}
\DeclareMathOperator{\Val}{Val}
\DeclareMathOperator{\Sat}{Sat}

\newcommand{\ordmin}{\funcname{min}}
\newcommand{\ordmax}{\funcname{max}}
\newcommand{\Translate}{\funcname{E}}

\newcommand{\Double}{\funcname{Double}}

\newcommand{\First}{\funcname{First}}
\newcommand{\Last}{\funcname{Last}}
\newcommand{\Succ}{\funcname{Succ}}
\newcommand{\FirstTuple}{\funcname{FirstTuple}}
\newcommand{\LastTuple}{\funcname{LastTuple}}
\newcommand{\LexPred}{\funcname{LexPred}}
\newcommand{\LexSucc}{\funcname{LexSucc}}
\newcommand{\Next}{\funcname{Next}}
\newcommand{\Tag}{\funcname{Tag}}
\newcommand{\Data}{\funcname{Data}}
\newcommand{\Char}{\funcname{Char}}
\newcommand{\InputLast}{\funcname{InputLast}}

\newcommand{\All}{\funcname{All}}
\newcommand{\AllAbove}{\funcname{AllAbove}}
\newcommand{\AllBelow}{\funcname{AllBelow}}
\newcommand{\Some}{\funcname{Some}}
\newcommand{\SomeBelow}{\funcname{SomeBelow}}
\newcommand{\At}{\funcname{At}}
\newcommand{\Zero}{\funcname{Zero}}
\newcommand{\Max}{\funcname{Max}}
\newcommand{\Nonzero}{\funcname{Nonzero}}
\newcommand{\Nonmax}{\funcname{Nonmax}}
\newcommand{\Eq}{\funcname{Eq}}
\newcommand{\Less}{\funcname{Less}}
\newcommand{\Same}{\funcname{Same}}
\newcommand{\Pred}{\funcname{Pred}}
\newcommand{\SPred}{\funcname{SPred}}
\newcommand{\SSucc}{\funcname{SSucc}}
\newcommand{\Code}{\funcname{Code}}

\newcommand{\Input}{\funcname{Input}}
\newcommand{\AfterInput}{\funcname{AfterInput}}
\newcommand{\Init}{\funcname{Init}}
\newcommand{\Cell}{\funcname{Cell}}
\newcommand{\Call}{\funcname{Call}}

\newcommand{\Prune}{\funcname{Prune}}
\newcommand{\Occ}{\funcname{Occ}}
\newcommand{\Row}{\funcname{Row}}
\newcommand{\Scan}{\funcname{Scan}}
\newcommand{\Use}{\funcname{Use}}

\newcommand{\TupleAll}{\funcname{TupleAll}}
\newcommand{\TupleAbove}{\funcname{TupleAbove}}
\newcommand{\TupleBelow}{\funcname{TupleBelow}}
\newcommand{\TupleTail}{\funcname{TupleTail}}
\newcommand{\VZero}{\funcname{VZero}}
\newcommand{\VMax}{\funcname{VMax}}
\newcommand{\VNonzero}{\funcname{VNonzero}}
\newcommand{\VNonmax}{\funcname{VNonmax}}
\newcommand{\VSame}{\funcname{VSame}}
\newcommand{\VEq}{\funcname{VEq}}
\newcommand{\VLess}{\funcname{VLess}}
\newcommand{\VSucc}{\funcname{VSucc}}
\newcommand{\VPred}{\funcname{VPred}}
\newcommand{\VSPred}{\funcname{VSPred}}
\newcommand{\VSSucc}{\funcname{VSSucc}}
\newcommand{\VAll}{\funcname{VAll}}
\newcommand{\VAllBelow}{\funcname{VAllBelow}}
\newcommand{\VSome}{\funcname{VSome}}
\newcommand{\VSomeAbove}{\funcname{VSomeAbove}}
\newcommand{\VSomeBelow}{\funcname{VSomeBelow}}
\newcommand{\VCode}{\funcname{VCode}}

\crefname{section}{Section}{Sections}
\Crefname{section}{Section}{Sections}
\crefname{subsection}{Section}{Sections}
\Crefname{subsection}{Section}{Sections}
\crefname{appendix}{Appendix}{Appendices}
\Crefname{appendix}{Appendix}{Appendices}
\crefname{theorem}{Theorem}{Theorems}
\Crefname{theorem}{Theorem}{Theorems}
\crefname{lemma}{Lemma}{Lemmas}
\Crefname{lemma}{Lemma}{Lemmas}
\crefname{proposition}{Proposition}{Propositions}
\Crefname{proposition}{Proposition}{Propositions}
\crefname{corollary}{Corollary}{Corollaries}
\Crefname{corollary}{Corollary}{Corollaries}
\crefname{definition}{Definition}{Definitions}
\Crefname{definition}{Definition}{Definitions}
\crefname{example}{Example}{Examples}
\Crefname{example}{Example}{Examples}
\crefname{claim}{Claim}{Claims}
\Crefname{claim}{Claim}{Claims}
\crefname{observation}{Observation}{Observations}
\Crefname{observation}{Observation}{Observations}

\newenvironment{myparagraph}[1]{%
  \par\addvspace{0.5\baselineskip}%
  \noindent\textbf{#1.}\nobreakspace\ignorespaces
}{\par\ignorespacesafterend}

\title{Higher-Order Generation of First-Order Theories}

\author{Ziyue Jin}{Peking University, Beijing, China}
  {zyjin@stu.pku.edu.cn}{https://orcid.org/0009-0005-2632-9423}{}
\author{Umang Mathur}{National University of Singapore, Singapore}
  {umathur@nus.edu.sg}{https://orcid.org/0000-0002-7610-0660}{}
\authorrunning{Z. Jin and U. Mathur}
\Copyright{Ziyue Jin and Umang Mathur}

\AtBeginDocument{\hypersetup{pdfsubject={Preprint}}}

\ccsdesc[500]{Theory of computation~Complexity theory and logic}
\ccsdesc[500]{Theory of computation~Finite Model Theory}
\ccsdesc[300]{Theory of computation~Automated reasoning}
\keywords{Higher-order recursion schemes, first-order theories, descriptive complexity, model checking, satisfiability}

\begin{document}

\maketitle


\begin{abstract}
We introduce a logic whose formulae are finite descriptions of possibly infinite first-order theories.
A theory is a set of first-order sentences, generated in our setting by a higher-order recursion scheme.
Such a scheme is a finite grammar whose parameters may denote trees or operations on trees.
We consider two interpretations: a model must satisfy every generated sentence under the conjunctive interpretation, and at least one under the disjunctive interpretation.
For schemes of order $k=0$, these theories coincide with the regular first-order theories recently introduced by Mathur et al.~\cite{Mathur2025Theories}.
We generalize this representation by allowing higher-order schemes, while retaining first-order quantification in every generated sentence.

We study the expressive power, model checking, and satisfiability of this logic.
Over finite ordered structures, we show that disjunctive theories of existential sentences generated at order at most $k$ capture $k$-$\EXPTIME$, for every $k\geq0$.
Allowing unrestricted first-order sentences captures $(k+1)$-$\EXPTIME$ at the same scheme order, where $0$-$\EXPTIME$ denotes polynomial time.
We also give model-checking algorithms in the corresponding time classes for each fixed signature, variable-name set, and order bound.
We then ask whether a theory has a model under either interpretation, and extend several decidability results for regular theories to every order.
However, we show that conjunctive satisfiability of quantifier-free theories, decidable at order zero, becomes undecidable already at order one.
Finally, we recover decidable conjunctive satisfiability at every order by separating higher-order generation of Boolean structure from independent regular generation of ground atoms.
Together, these results show how higher-order generation of first-order syntax captures the exponential-time hierarchy while retaining decidable satisfiability in restricted fragments.
\end{abstract}


\section{Introduction}\label{sec:intro}

The classical \emph{Entscheidungsproblem} asks for an algorithm to decide whether a first-order sentence is valid.
Its undecidability led to the study of decidable fragments of first-order logic~\cite{BorgerGradelGurevich1997}.
Deciding validity and satisfiability in such fragments remains central to program verification, database querying, and constraint solving.
Descriptive complexity relates the properties we can express in a logic to the computational resources needed to decide them~\cite{Immerman1999Descriptive}.

Families of first-order formulae arise naturally in program verification.
In the ``executions as formulae'' view, we represent each finite candidate execution by a formula describing its assignments, guards, and assumptions.
The formula is satisfiable if and only if the execution is feasible under the intended interpretation of the data operations.
To find a bug, we ask whether the formula for some execution reaching an error is satisfiable.
Loops and recursion can give rise to infinitely many such formulae, so we need a finite representation of the family.
Mathur et al.~\cite{Mathur2025Theories} introduced \emph{regular first-order theories}: sets of formulae whose syntax trees are recognized by a finite tree automaton.

We generalize regular first-order theories by using \emph{higher-order recursion schemes} to generate the formulae.
A recursion scheme is a finite typed grammar whose rules recursively build trees.
Its rules may take trees or operations on trees as parameters, much as a higher-order procedure can take another procedure as an argument~\cite{Ong2006ModelChecking}.
We use these schemes to generate syntax trees of first-order sentences and call the resulting set of sentences a \emph{theory}.

A formula of our logic consists of a scheme together with a choice of interpretation.
If the scheme generates the sentences $\varphi_1,\varphi_2,\ldots$, we read it as either
\begin{align*}
  \varphi_1\lor\varphi_2\lor\cdots
  \qquad\text{or}\qquad
  \varphi_1\land\varphi_2\land\cdots.
\end{align*}
A structure satisfies the disjunction if some generated sentence holds in it, and the conjunction if every generated sentence holds in it.
We study the expressive power, model checking, and satisfiability of the resulting fragment of infinitary first-order logic.

\begin{example}[Doubling a tree-building operation]\label{ex:intro-doubling}
  Let $c$ be a constant, $f$ a unary function symbol, and $P$ a unary predicate.
  Consider the grammar with start symbol $S$ and rules
  \begin{align*}
    S &\to A\,f,\\
    A\,h &\to \neg P(h\,c)
       \mid A\,(\Double\,h),\\
    \Double\,h\,x &\to h\,(h\,x).
  \end{align*}
  Here $\mid$ separates two alternative rules, and $h$ and $x$ are grammar parameters.
  The parameter $h$ is an operation on trees, and $\Double\,h$ is the operation that applies $h$ twice.
  The first rule for $A$ produces a sentence; the second replaces $h$ by $\Double\,h$ and continues recursively.
  Starting with $f$, the scheme therefore generates
  \begin{align*}
    \neg P(f(c)),\quad
    \neg P(f^2(c)),\quad
    \neg P(f^4(c)),\quad
    \neg P(f^8(c)),\quad\ldots,
  \end{align*}
  where $f^m(c)$ abbreviates $m$ nested applications of $f$ to $c$.
  The scheme is higher order because $A$ and $\Double$ take the operation $h$ as an argument.
  Under the disjunctive reading, $P$ must fail at $f^{2^n}(c)$ for some $n\geq0$; under the conjunctive reading, it must fail at every such iterate.
\end{example}

At order zero, rules have no parameters, and the generated theories are exactly regular first-order theories.
Order one permits tree parameters; order two permits parameters denoting operations on trees, as in \cref{ex:intro-doubling}.

We write $\bigvee_{k-\HORS}\bL$ and $\bigwedge_{k-\HORS}\bL$ for the disjunctive and conjunctive interpretations of schemes of order at most $k$ generating sentences in a first-order fragment $\bL$.
Each scheme uses a finite signature and a fixed finite set of variable names, which may be reused in different scopes.
We consider unrestricted first-order sentences ($\FO$), as well as the existential and universal fragments ($\exists\FO$ and $\forall\FO$).
In these two fragments, negation is restricted to atoms, and all quantifiers are existential or all are universal, respectively.

\begin{myparagraph}{Our contributions}
\begin{enumerate}
  \item \textbf{A logic and its basic properties.}
  We define the logic over finite and infinite models.
  We show how to negate a formula by negating its generated sentences and exchanging the two interpretations, without increasing scheme order (\cref{prop:logic-duality}).
  The finite set of variable names imposes an expressive limitation at every order: we cannot define even cardinality on finite models with equality alone (\cref{prop:logic-parity}).
  Over infinite models, we show that our logic is strictly less expressive than full second-order logic (\cref{prop:logic-infinite-so}).

  \item \textbf{Descriptive complexity and model checking.}
  Over finite ordered models, for every $k\geq0$, we prove (\cref{thm:descriptive-existential,thm:descriptive-first-order}):
  \begin{align*}
    \bigvee_{k-\HORS}\exists\FO
      &\quad\text{captures}\quad k\text{-}\EXPTIME,\\
    \bigvee_{k-\HORS}\FO
      &\quad\text{captures}\quad (k+1)\text{-}\EXPTIME.
  \end{align*}
  Here $0$-$\EXPTIME$ denotes polynomial time.
  The input vocabulary includes the order, its endpoints, and its successor relation.
  Each query is defined by a fixed presentation, independent of the input model.
  The same classes are captured by $\bigwedge_{k-\HORS}\forall\FO$ and $\bigwedge_{k-\HORS}\FO$, respectively.
  For each fixed signature, variable-name set, and order bound, we give model-checking algorithms in the corresponding time classes, measured in the combined size of the scheme and model (\cref{thm:model-checking-existential,thm:model-checking-first-order}).

  \item \textbf{Satisfiability and its boundaries.}
  We ask whether some model, finite or infinite, satisfies a generated theory under the chosen interpretation.
  We extend the bounded-model method for regular theories~\cite{Mathur2025Theories} to every scheme order by combining bounds on model size with our model-checking algorithms.
  Using this method, we show that satisfiability is decidable under both readings for purely existential and purely universal relational sentences, allowing constants and equality (\cref{thm:sat-relational}).
  Disjunctive satisfiability also remains decidable at every order for theories of coherent existential sentences, where coherence restricts when values can be forgotten and recomputed (\cref{thm:sat-coherent}).

  Conjunctive satisfiability of quantifier-free theories is decidable at order zero.
  At every order $k\geq1$, however, we show that both conjunctive and disjunctive satisfiability are undecidable already for quantifier-free sentences with equality, constants, and two unary function symbols (\cref{thm:sat-qf-unary}).
  We recover decidable conjunctive satisfiability at every order by allowing the higher-order scheme to generate the Boolean structure while separate regular grammars independently generate the ground atomic formulae (\cref{thm:sat-regular-atoms}).
\end{enumerate}
\end{myparagraph}

Over finite ordered structures, classical descriptive-complexity results capture polynomial time by first-order logic with least fixed points, and the exponential-time hierarchy by higher-order fixed-point logics~\cite{Immerman1999Descriptive,FreireMartins2011}.
We capture the same complexity classes through infinitary combinations of first-order sentences (\cref{cor:logic-comparison}).
Increasing the order of the finite generator gives us this expressive power while all quantification over the input structure remains first order.

\begin{myparagraph}{Organization}
After the preliminaries in \cref{sec:preliminaries}, we define the logics in \cref{sec:logic}, establish model-checking and descriptive-complexity results in \cref{sec:model-checking,sec:descriptive-complexity}, and study satisfiability in \cref{sec:satisfiability}.
Detailed proofs and constructions appear in the appendices.
\end{myparagraph}


\section{Preliminaries}\label{sec:preliminaries}

We recall the notation for first-order logic~\cite{Immerman1999Descriptive} and higher-order recursion schemes.

\subsection{First-order logic}

\begin{myparagraph}{Syntax and semantics}
A \emph{first-order signature} $\sigma = \tuple{\cC, \cR, \cF}$ consists of finite, pairwise disjoint sets of constant, relation, and positive-arity function symbols.
Each relation or function symbol $a$ has a fixed \emph{arity} $\rk(a)$.
We fix a finite set $\cV$ of variable names.

We write $\FO^{\sigma,\cV}$, or simply $\FO$, for the set of first-order formulae with equality over $(\sigma,\cV)$.
A \emph{sentence} is a formula with no free variables.

A model $\cM$ has a nonempty universe $U$ and interprets each function symbol as a total function.
Unless stated otherwise, the universe may be finite or infinite.
For an \emph{assignment} $\nu:\cV\to U$, we use the usual satisfaction relation $\cM,\nu\vDash\varphi$ and omit $\nu$ when $\varphi$ is a sentence.
A formula is \emph{satisfiable} if it holds in some model under some assignment.

\end{myparagraph}

\begin{myparagraph}{Syntactic fragments}
A formula is in \emph{negation normal form} (NNF) if negation occurs only immediately before atomic formulae.
We write $\exists\FO$ and $\forall\FO$ for the NNF fragments containing only existential and only universal quantifiers, respectively.

\end{myparagraph}

\subsection{Higher-order recursion schemes}

\begin{myparagraph}{Ranked alphabet}
A \emph{ranked alphabet} is a finite set $\Sigma$ equipped with a \emph{rank function} $\rk:\Sigma\to\bN$.
We write $a^r$ to indicate that $\rk(a)=r$, and $T_\Sigma$ for the set of finite ordered trees over $\Sigma$ in which every node labelled by $a$ has exactly $\rk(a)$ children.

\end{myparagraph}

\begin{myparagraph}{Higher-order recursion scheme}
\emph{Simple types} are generated by $\tau ::= o \mid \tau\to\tau$, where $o$ is the type of trees and arrows associate to the right.
We define the \emph{order} of a type by $\ord(o)=0$ and $\ord(\tau\to\rho)=\max\{\ord(\tau)+1,\ord(\rho)\}$.

A \emph{nondeterministic higher-order recursion scheme} is a tuple $G=\tuple{\Sigma,\cN,\cP,S}$, where $\Sigma$ is a ranked alphabet, $\cN$ is a finite set of typed nonterminals, $\cP$ is a finite set of rules, and $S\in\cN$ is a start nonterminal of type $o$.
A terminal of rank $r$ has type $o\to\cdots\to o\to o$ with $r$ arguments.

\emph{Applicative expressions} are given by $M ::= u \mid F \mid a \mid M\,M$, where $u$ is a typed parameter, $F\in\cN$, and $a\in\Sigma$.
Application associates to the left and follows the usual typing rule: if $M:\tau\to\rho$ and $N:\tau$, then $M\,N:\rho$.
A \emph{rule} has the form $F\,u_1\cdots u_r\to M$, where $F:\tau_1\to\cdots\to\tau_r\to o$ and the parameters $u_i:\tau_i$ are distinct.
The right-hand side $M$ has type $o$ and no free parameters other than $u_1,\ldots,u_r$.
A nonterminal may have several rules.

Each rule $F\,u_1\cdots u_r\to M$ induces a \emph{rewriting step}
\[
  F\,M_1\cdots M_r\Rightarrow_G M[M_1/u_1,\ldots,M_r/u_r],
\]
which may be applied in any well-typed context.
Arguments need not be unfolded before substitution.
If substitution creates several copies of an argument, each copy may subsequently be unfolded using different rules.

The \emph{finite tree language generated by $G$} is $\cL(G)=\setpred{t\in T_\Sigma}{S\Rightarrow_G^* t}$.
Only finite derivations ending in terminal trees contribute to $\cL(G)$.

The \emph{order of $G$} is the maximum order of its nonterminal types; terminal types are not counted.
We call a scheme of order at most $k$ a $k-\HORS$.
In particular, $0-\HORS$ are regular tree grammars and generate exactly the regular tree languages.
\end{myparagraph}

\section{Logic}\label{sec:logic}

\subsection{Generated first-order logics}

We use higher-order recursion schemes to give finite descriptions of possibly infinite sets of first-order sentences.
The order refers to the scheme that generates the syntax; quantification within each generated sentence remains first order.

\begin{definition}[First-order alphabet]
  For a finite first-order signature $\sigma = \tuple{\cC, \cR, \cF}$ and a finite set $\cV$ of variable names, we define the ranked alphabet
  \begin{align*}
    \Sigma_\FO^{\sigma, \cV} &= \{\bot^0, \top^0, \neg^1, =^2, \lor^2, \land^2\} \cup \{(\exists x)^1, (\forall x)^1\}_{x \in \cV}\\
                             &\cup \{x^0\}_{x \in \cV} \cup \{c^0\}_{c \in \cC} \cup \{R^{\rk(R)}\}_{R \in \cR} \cup \{f^{\rk(f)}\}_{f \in \cF}.
  \end{align*}
  We omit $\sigma$ and $\cV$ when clear from context.
\end{definition}

We identify first-order terms and formulae with their syntax trees over $\Sigma_\FO$.
Both terms and formulae have ground type $o$, so typing alone does not ensure that generated trees are sentences.
We therefore require explicitly that every generated tree be a sentence in the chosen fragment.

\begin{definition}[Generated first-order logic]\label{def:generated-logic}
  Fix a finite signature $\sigma$, a finite set $\cV$ of variable names, $k \in \bN$, a fragment $\bL \subseteq \FO^{\sigma,\cV}$, and $\bigodot \in \set{\bigvee, \bigwedge}$.
  A formula of $\bigodot_{k-\HORS}\bL$ is a finite presentation $\bigodot G$, where $G$ is a $k-\HORS$ over $\Sigma_\FO$ and every tree in $\cL(G)$ is a sentence in $\bL$.
  For any first-order model $\cM$, define
  \begin{align*}
    \cM \vDash \bigvee G &\iff \exists \varphi \in \cL(G).~\cM \vDash \varphi
    \qquad \text{and} \qquad
    \cM \vDash \bigwedge G &\iff \forall \varphi \in \cL(G).~\cM \vDash \varphi.
  \end{align*}
\end{definition}

We evaluate all generated sentences in the same model $\cM$, which may be finite or infinite.
If $\cL(G)=\emptyset$, then $\bigvee G$ is false and $\bigwedge G$ is true.
Only completed derivations of finite sentences contribute to either interpretation.

First-order variable names $x\in\cV$ are terminal symbols and are not renamed when expressions are substituted for grammar parameters.
A quantifier can therefore bind occurrences of $x$ introduced into its scope by grammar substitution.
For example, the rules $Q\,u\to(\exists x)\,u$ and $S\to Q\,(x=c)$ generate the sentence $\exists x.~x=c$.

\begin{example}[Choosing one generated sentence]\label{ex:uniform-choice}
  Consider the order-$0$ scheme with rules $S\to(\forall x)\,A$, $A\to P\,x$, and $A\to Q\,x$.
  Its disjunctive reading is $(\forall x.~P(x))\lor(\forall x.~Q(x))$.
  In a two-element model where $P$ holds only at the first element and $Q$ only at the second, neither generated sentence holds.
  The disjunctive presentation is therefore false, although $\forall x.~(P(x)\lor Q(x))$ is true.
  The choice of generated sentence is made before its quantifiers are evaluated and cannot depend on the values assigned to its variables.
\end{example}

\subsection{Basic properties}

At order zero, we recover the disjunctive and conjunctive readings of regular first-order theories~\cite{Mathur2025Theories}, since order-$0$ schemes generate exactly the regular tree languages.
In particular, each first-order sentence $\varphi$ has an order-$0$ singleton presentation $S\to\varphi$ under either reading.
For each fragment $\bL$ and reading $\bigodot$, we have $\bigodot_{k-\HORS}\bL \subseteq \bigodot_{(k+1)-\HORS}\bL$.

Negating the generated sentences and exchanging the two readings gives the negation of a presentation.
\begin{restatable}[Boolean duality]{proposition}{booleanDuality}\label{prop:logic-duality}
  For every $k$-$\HORS$ $G$ generating first-order sentences, we can effectively construct a $k$-$\HORS$ $G^\neg$ such that, over all models, $\neg\bigvee G \equiv \bigwedge G^\neg$ and $\neg\bigwedge G \equiv \bigvee G^\neg$.
  If $G$ generates $\exists\FO$ sentences, then $G^\neg$ can be chosen to generate $\forall\FO$ sentences, and conversely.
  For each fixed $k$, the construction takes polynomial time.
\end{restatable}

We give the construction and proof in \cref{app:logic-duality}.

\subsection{Finite-variable bounds}

To bound expressive power, we use the finite set of variable names to place our logics within finite-variable infinitary logic.

The infinitary logic $L_{\omega_1\omega}$ extends first-order logic with countable conjunctions and disjunctions.
Its fragment $L_{\omega_1\omega}^m$ uses at most $m$ variable names throughout each formula, including all its infinitary constituents.
We write $L_{\omega_1\omega}^{\omega}=\bigcup_{m<\omega}L_{\omega_1\omega}^m$.
These infinitary formulae need not have effective descriptions.

\begin{proposition}[Infinitary interpretation]\label{prop:logic-invariance}
  Over all models, $\bigodot_{k-\HORS}\FO \subseteq L_{\omega_1\omega}^{\omega}$ for every $k\in\bN$ and $\bigodot\in\set{\bigvee,\bigwedge}$.
  More precisely, a presentation using $m$ variable names has an interpretation in $L_{\omega_1\omega}^m$.
\end{proposition}

\begin{proof}
  Let $\bigodot G$ be a presentation using $m$ variable names.
  The language $\cL(G)$ is countable, since it consists of finite trees over a finite alphabet.
  Every sentence in $\cL(G)$ uses the same set of at most $m$ variable names.
  Thus $\bigodot_{\varphi\in\cL(G)}\varphi$ belongs to $L_{\omega_1\omega}^m$ and has the required semantics.
\end{proof}

Our logics inherit the expressive limitations of $L_{\omega_1\omega}^{\omega}$, including the following parity obstruction~\cite[Example~2.22]{KolaitisVardi1992Infinitary}.

\begin{proposition}[Parity]\label{prop:logic-parity}
  Over the empty nonlogical signature with equality, no presentation in $\bigodot_{k-\HORS}\FO$ defines even cardinality on finite models, for any $k\in\bN$ and $\bigodot\in\set{\bigvee,\bigwedge}$.
\end{proposition}

Over infinite models, we show that our logics are strictly less expressive than full second-order logic; we give this comparison and further expressive limitations in \cref{app:logic-infinite-models,app:logic-infinite-so}.

\section{Model checking}\label{sec:model-checking}

Throughout this section, we fix the finite signature $\sigma$, finite set $\cV$ of variable names, and order bound $k$.
For a fragment $\bL\subseteq\FO$ and $\bigodot\in\set{\bigvee,\bigwedge}$, the model-checking problem takes as input a finite $\sigma$-model $\cM$ and a $k$-$\HORS$ $G$ generating sentences in $\bL$, and asks whether $\cM\vDash\bigodot G$.
The model is represented explicitly by its universe and the tables interpreting its symbols.
We measure complexity in the combined size $\size{G}+\size{\cM}$ of the scheme and model.
As usual, $0-\EXPTIME$ denotes polynomial time.

We first consider the disjunctive reading.
Our algorithm constructs a tree automaton recognizing the sentences that hold in $\cM$, then checks whether $G$ generates any sentence accepted by this automaton.
The construction relies on the fact that the first-order sentences over $(\sigma,\cV)$ that are true in a fixed finite model form a regular tree language~\cite[Lemma~4.2]{Mathur2025Theories}.
We use nondeterministic bottom-up tree automata that evaluate formulae under assignments~\cite[Section~5.1]{KrogmeierMadhusudan2022Learning}, counting both states and transitions in their size.
The constants hidden in the next two bounds depend only on $\sigma$ and $\cV$; we give the constructions and proofs in \cref{app:model-checking}.

\begin{restatable}{lemma}{existentialAutomaton}\label{lem:efo-theory-reg}
  For every finite $\sigma$-model $\cM$, the $\exists\FO$ sentences over $(\sigma,\cV)$ that are true in $\cM$ are recognized by a tree automaton of size $\size{\cM}^{O(1)}$.
  Such an automaton can be constructed in time $\size{\cM}^{O(1)}$.
\end{restatable}

\begin{restatable}{lemma}{firstOrderAutomaton}\label{lem:fo-theory-reg}
  For every finite $\sigma$-model $\cM$, the $\FO$ sentences over $(\sigma,\cV)$ that are true in $\cM$ are recognized by a tree automaton of size $2^{\size{\cM}^{O(1)}}$.
  Such an automaton can be constructed in time $2^{\size{\cM}^{O(1)}}$.
\end{restatable}

It remains to check whether the scheme generates a tree accepted by the automaton.
We obtain the following bound from standard results on higher-order model checking~\cite{Ong2006ModelChecking,Parys2021ModelChecking}.
We give the reduction for the finite tree languages considered here in \cref{app:hors-automaton-nonemptiness}.

\begin{lemma}\label{lem:hors-automaton-nonemptiness}
  Fix $k\in\bN$.
  Given a $k-\HORS$ $G$ and a finite tree automaton $A$ over the same ranked alphabet, nonemptiness of $\cL(G)\cap\cL(A)$ can be decided in time $\exp_k\bigl((\size{G}+\size{A})^{O(1)}\bigr)$, where $\exp_0(n)=n$ and $\exp_{i+1}(n)=2^{\exp_i(n)}$.
\end{lemma}

Combining these bounds gives the following results; the automaton for unrestricted first-order logic contributes one additional exponential.

\begin{theorem}\label{thm:model-checking-existential}
  Model checking for $\bigvee_{k-\HORS}\exists\FO$ and $\bigwedge_{k-\HORS}\forall\FO$ is in $k-\EXPTIME$.
\end{theorem}

\begin{theorem}\label{thm:model-checking-first-order}
  Model checking for $\bigvee_{k-\HORS}\FO$ and $\bigwedge_{k-\HORS}\FO$ is in $(k+1)-\EXPTIME$.
\end{theorem}

\begin{proof}[Proof of both theorems]
  By \cref{lem:efo-theory-reg,lem:fo-theory-reg}, we can construct a tree automaton $A_\cM$ recognizing exactly the sentences of the chosen fragment that are true in $\cM$.
  Then $\cM\vDash\bigvee G$ if and only if $\cL(G)\cap\cL(A_\cM)\neq\emptyset$.
  Applying \cref{lem:hors-automaton-nonemptiness} gives the $k-\EXPTIME$ and $(k+1)-\EXPTIME$ bounds for $\exists\FO$ and full $\FO$, respectively.
  We give the complexity calculation in \cref{app:model-checking-complexity}.
  The conjunctive cases follow from \cref{prop:logic-duality} and closure of these complexity classes under complement.
\end{proof}

The same upper bounds hold for data complexity, where the scheme is fixed and only the model is part of the input.
In \cref{sec:descriptive-complexity}, we use these bounds to characterize the queries definable at each scheme order.

\section{Descriptive complexity}\label{sec:descriptive-complexity}

\subsection{Ordered models}

We consider Boolean queries, viewed as classes of finite models closed under isomorphism.

The parity obstruction in \cref{prop:logic-parity} rules out capturing polynomial time over arbitrary finite models, so we work over ordered models as in classical descriptive complexity~\cite{Immerman1999Descriptive}.

A finite model is \emph{ordered} if its signature contains a distinguished strict linear order $<$, constants $\ordmin$ and $\ordmax$ naming its endpoints, and a binary relation $\Succ$ interpreted as the immediate-successor relation of $<$.
Although the endpoints and successor are determined by $<$, we include them in the input vocabulary because the existential fragment uses them directly.
For full $\FO$, we can eliminate the extra symbols using fixed first-order definitions.
We abbreviate $x=\ordmin$ and $x=\ordmax$ by $\First(x)$ and $\Last(x)$, respectively.

For each fixed signature, query complexity is measured on the canonical encoding of an ordered model by its universe and symbol tables.
A logic \emph{captures} a complexity class over finite ordered models if every fixed sentence defines a query in the class, and every query in the class is defined by a fixed sentence of the logic.
For our logics, this means that the scheme and its finite set of variable names may depend on the query but not on the input model.
We evaluate the generated sentences on the original ordered model.

We use the tower functions $\exp_i$ from \cref{lem:hors-automaton-nonemptiness}, so $k-\EXPTIME$ is the union of $\DTIME(\exp_k(n^c))$ over constants $c\geq1$.
In particular, the case $k=0$ is polynomial time.

\subsection{Results}

\begin{theorem}\label{thm:descriptive-existential}
  For $k \in \bN$, both $\bigvee_{k-\HORS} \exists\FO$ and $\bigwedge_{k-\HORS} \forall\FO$ capture $k-\EXPTIME$ over finite ordered models.
\end{theorem}

\begin{proof}
  The upper bound follows from \cref{thm:model-checking-existential} with the scheme fixed.
  For the converse, fix a deterministic machine deciding the query on encoded models in $k$-$\EXPTIME$.
  We sketch its simulation here and give the details in \cref{prop:descriptive-lower-bounds}.
  We first work over a padded ordered word model with domain $D$ of size $N$.
  Adapting complemented higher-order counters~\cite{CharalambidisNomikosRondogiannis2019HODatalog}, we construct counter domains $I_0=D$ and $I_{i+1}=\set{0,1}^{I_i}$, so $\size{I_i}=\exp_i(N)$.
  We represent time and tape positions in $I_k$.
  To check an entire address interval, we use scanning rules that generate one conjunct for each address, using only existential first-order binders.
  To verify the simulation, we show that $\Cell_\gamma(P,T)$ has a true finite unfolding if and only if position $P$ contains symbol $\gamma$ at time $T$.
  The proof proceeds by induction on time: the recursive premises check the cell and its two neighbours at the preceding time.
  After making halting states stationary, we express acceptance by testing for an accepting cell at the final simulated time.
  The resulting scheme has order at most $k$.
  Both the scheme and its finite set of variable names depend only on the machine.

  To obtain a presentation over the original input model, we represent the padded word by a fixed tuple interpretation.
  This translation preserves both the existential fragment and the scheme order.
  In \cref{prop:descriptive-lower-bounds}, we choose the required padding and handle singleton models separately.
  We derive the conjunctive characterization using closure of $k-\EXPTIME$ under complement and \cref{prop:logic-duality}.
\end{proof}

\begin{theorem}\label{thm:descriptive-first-order}
  For $k \in \bN$, both $\bigvee_{k-\HORS} \FO$ and $\bigwedge_{k-\HORS} \FO$ capture $(k+1)-\EXPTIME$ over finite ordered models.
\end{theorem}

\begin{proof}
  The upper bound follows from \cref{thm:model-checking-first-order} with the scheme fixed.
  For the converse, we enlarge the base address domain of the preceding simulation to obtain one additional exponential without increasing the scheme order.

  Over a padded word domain $D$ of size $N\geq2$, we use relations of a fixed arity $r$ as base addresses: take $I_0=\powerset{D^r}$ and $I_{i+1}=\set{0,1}^{I_i}$, so $\size{I_i}=\exp_{i+1}(N^r)$.
  We store the base addresses in relation variables and retain the types of the higher-level counter components.
  At this stage, the simulation generates sentences with existential first-order and existential relation quantifiers, using fixed finite sets of names and literal templates.

  We then eliminate the relation quantifiers.
  First, we replace each disjunction in the intermediate sentences by a choice in the grammar.
  Since all quantifiers in these sentences are existential, this preserves the disjunctive interpretation (\cref{lem:descriptive-pruning}).
  Each relation-literal occurrence independently generates a table of bits, with one bit for each tuple in $D^r$.
  First-order successor and endpoint guards enforce that every tuple occurs in exactly one row.
  Occurrences bound by the same relation quantifier must use the same table; otherwise, independent choices could incorrectly satisfy $P(x)\land\neg P(x)$.

  To enforce agreement, at each relation binder we introduce a shared challenge tuple $\bar i_P$ and response $p_P$.
  Every literal occurrence bound by that quantifier reports its table's bit at the challenged address.
  At its own argument tuple, a positive occurrence requires bit $1$ and a negative occurrence requires bit $0$.
  Writing $B_j$ for the generated tables, $\beta_0(p)=\First(p)$, and $\beta_1(p)=\Last(p)$, the synchronization condition is, in semantic notation,
  \begin{align*}
    \forall\bar i_P.~\exists p_P.~\bigwedge_j\beta_{B_j(\bar i_P)}(p_P).
  \end{align*}
  The generated syntax fixes all tables before the challenge is evaluated.
  A common response therefore exists at every address if and only if the tables agree pointwise.
  Choosing disjuncts in the grammar prevents the checked branch from changing with the challenge.
  We formalize this argument, including nested bindings, in \cref{lem:descriptive-relation-synchronization}.

  The nonterminals generating the tables have ground type, and the added quantifiers belong to the generated first-order syntax.
  The elimination therefore preserves scheme order, including at $k=0$ (\cref{prop:descriptive-relation-elimination}).
  We verify the lifted simulation in \cref{app:descriptive-lift}, then return to the original input model using the same tuple interpretation.
  Using closure under complement and \cref{prop:logic-duality}, we obtain the same expressiveness for $\bigwedge_{k-\HORS}\FO$.
\end{proof}

\section{Satisfiability}\label{sec:satisfiability}

For fixed $k\in\bN$, $\bL\subseteq\FO$, and $\bigodot\in\set{\bigvee,\bigwedge}$, the \emph{satisfiability problem} takes a $k$-$\HORS$ $G$ generating sentences in $\bL$ and asks whether $\cM\vDash\bigodot G$ for some finite or infinite model $\cM$.
Under the conjunctive reading, all generated sentences must hold in the same model.

\subsection{Transfer from regular theories}

Every regular tree grammar is a $k$-$\HORS$, so the undecidability results of Mathur et al.~\cite[Theorems~1.2 and~1.3]{Mathur2025Theories} transfer to every order, with free source parameters replaced by fresh constants.
For example, satisfiability is undecidable under both readings even for existential sentences using only equality, constants, and two unary function symbols; see \cref{app:sat-inherited-undecidability} for the precise fragments.

The bounded-model arguments of Mathur et al.~\cite[Sections~4 and~7]{Mathur2025Theories} use finite-model checking, while their coherence argument tests whether the generated language intersects a regular tree language.
Both tests are decidable for our higher-order presentations by \cref{sec:model-checking}, so these arguments extend to every scheme order.
We give the full proofs in \cref{app:sat-bounded-models,app:sat-coherence}.

\begin{definition}[Bounded model properties]\label{def:bounded-model-properties}
  A fragment $\bL$ has the \emph{weak bounded model property} if, for every finite signature $\sigma$ and finite set $\cV$ of variable names, there is a finite bound $b(\sigma,\cV)$ such that every satisfiable sentence in $\bL$ over $(\sigma,\cV)$ has a model of size at most $b(\sigma,\cV)$.
  It has the \emph{strong bounded model property} if this bound is computable from $(\sigma,\cV)$.
  Write $\bL^\land$ for the finite conjunctions of sentences in $\bL$.
\end{definition}

\begin{restatable}{theorem}{boundedModels}\label{thm:sat-bounded-models}
  Let $k\in\bN$ and $\bL\subseteq\FO$.
  If $\bL^\land$ has the weak bounded model property, then satisfiability for $\bigwedge_{k-\HORS}\bL$ is decidable.
  If $\bL$ has the strong bounded model property, then satisfiability for $\bigvee_{k-\HORS}\bL$ is decidable.
\end{restatable}

The two fragments below are closed under finite conjunctions and have the strong bounded model property~\cite[Section~6.1]{Mathur2025Theories}, so \cref{thm:sat-bounded-models} applies.

\begin{theorem}\label{thm:sat-relational}
  Over relational signatures allowing constants and equality but no positive-arity functions, satisfiability is decidable for both $\bigodot_{k-\HORS}\exists\FO$ and $\bigodot_{k-\HORS}\forall\FO$, for every $k\in\bN$ and $\bigodot\in\set{\bigvee,\bigwedge}$.
\end{theorem}

\begin{myparagraph}{Coherent formulae}
Let $\Coh$ be the class of coherent existential sentences over shallow terms, in the value-tracking formulation of Mathur et al.~\cite[Section~7 and Appendix~C]{Mathur2025Theories}.
Coherence prevents a tracked value from losing every variable representative and later acquiring one again along a path.
We define coherence precisely in \cref{app:sat-coherence}.

\end{myparagraph}

\begin{restatable}{theorem}{coherentSatisfiability}\label{thm:sat-coherent}
  For every $k\in\bN$, satisfiability for $\bigvee_{k-\HORS}\Coh$ is decidable.
\end{restatable}

\subsection{The quantifier-free boundary}\label{sec:sat-qf-boundary}

Conjunctive satisfiability of regular quantifier-free theories with equality and uninterpreted functions (EUF) is decidable~\cite[Theorem~1.5]{Mathur2025Theories}.
This result does not extend to unrestricted higher-order presentations.

\begin{restatable}{theorem}{quantifierFreeUndecidability}\label{thm:sat-qf-unary}
  For every $k\geq1$, both conjunctive and disjunctive satisfiability are undecidable for $k$-$\HORS$ generating ground quantifier-free sentences.
  This holds over a fixed signature consisting of two unary function symbols and finitely many constants, with equality as the only predicate.
\end{restatable}

\begin{proof}[Proof sketch]
  We use the undecidability of intersection emptiness for two context-free word languages.
  For the conjunctive reading, we generate disequalities between terms encoding words from the two languages.
  For the disjunctive reading, we generate ground constraints describing two words along the same function orbit, using distinct markers for letters and for the end of a word.
  The conjunctive presentation is satisfiable if and only if the languages are disjoint; the disjunctive presentation is satisfiable if and only if they intersect.
  We give the order-$1$ constructions in \cref{app:sat-qf-undecidable}.
\end{proof}

For $k\geq1$, this eliminates all first-order quantifiers from the inherited unary-function lower bound.

\subsection{Regular atomic components}\label{sec:sat-regular-components}

We recover decidability by letting the higher-order scheme generate Boolean structure while separate regular grammars generate its atomic formulae.
The restriction is on how the presentation is assembled: choices at distinct atomic occurrences are independent.

\begin{definition}[Regular atomic components]\label{def:hors-reg}
  A $k$-$\HORS/\REG$ presentation is a pair $(G,(A_i)_{i\in I})$, where $I$ is finite.
  The scheme $G$ is a $k$-$\HORS$ over $\{\bot^0,\top^0,\neg^1,\land^2,\lor^2\}\cup\{a_i^0:i\in I\}$, and each $A_i$ is a nonempty regular tree language of ground atomic formulae, given by a finite tree automaton.
  Its generated language consists of the formulae obtained from any $t\in\cL(G)$ by independently replacing each occurrence of $a_i$ with an atom from $A_i$.
  We write $\bigodot_{k-\HORS/\REG}\bL$ for these presentations with all generated sentences in $\bL$, under the usual reading $\bigodot$.
\end{definition}

The atomic grammars receive no term parameters from $G$.
We obtain an ordinary $k$-$\HORS$ presentation by substituting their ground start symbols for the $a_i$.
If an input atomic component is empty, we first use a regular tree filter to remove all skeletons containing its interface symbol.

\begin{restatable}{theorem}{regularAtomicComponents}\label{thm:sat-regular-atoms}
  For every $k\in\bN$, conjunctive satisfiability of $k$-$\HORS/\REG$ presentations is decidable.
\end{restatable}

\begin{proof}[Proof sketch]
  For each atomic component, we choose one of three requirements: all its atoms must be true, all must be false, or their truth values are unrestricted.
  We check satisfiability of the resulting regular theory using~\cite[Theorem~1.5]{Mathur2025Theories}.
  We construct a finite tree automaton recognizing the Boolean skeletons that are true under every permitted, independent choice of truth values at their atomic occurrences.
  We then combine this automaton with the order-$k$ scheme to check that every generated skeleton has this property.
  The presentation is satisfiable if and only if some choice of requirements passes both tests; we give the full argument in \cref{app:sat-regular-atoms}.
\end{proof}


\section{Related work}\label{sec:related-work}

\begin{myparagraph}{Regular theories}
The closest work is that of Mathur et al.~\cite{Mathur2025Theories}, who study conjunctive and disjunctive satisfiability of first-order theories presented by finite tree automata.
These are the order-zero presentations of our logic.
We extend their bounded-model and coherence criteria to every scheme order.
Their decidability result for conjunctive quantifier-free theories, however, fails already at order one, even with only two unary function symbols.
We recover decidability by restricting atomic components to regular languages while allowing higher-order generation of the Boolean structure (\cref{thm:sat-qf-unary,thm:sat-regular-atoms}).
\end{myparagraph}

\begin{myparagraph}{Higher-order logics}
Classical descriptive complexity relates logical definability to computational complexity over finite structures~\cite{Immerman1999Descriptive}.
Over finite ordered structures, first-order logic with least fixed points, $\FO(\LFP)$, captures polynomial time. More generally, its order-$(k+1)$ extension $\HO^{k+1}(\LFP)$ captures $k$-$\EXPTIME$~\cite{FreireMartins2011,BruseKronenbergerLange2022}.
Our disjunctive presentations have the same expressive power using order-$k$ generation of existential first-order sentences, or order-$(k-1)$ generation of unrestricted first-order sentences when $k\geq1$ (see \cref{app:descriptive-fixed-point}).
Here the higher-order objects are operations on syntax; quantification in the generated sentences remains first order.
Bruse et al.~\cite{BruseKronenbergerLange2022} also show that order-$k$ polyadic higher-order fixpoint logic captures the bisimulation-invariant $k$-$\EXPTIME$ queries on finite transition systems.
We use their result to derive the inclusions for our bisimulation-invariant queries in \cref{app:descriptive-phfl}.
\end{myparagraph}

\begin{myparagraph}{Higher-order Datalog}
Charalambidis et al.~\cite{CharalambidisNomikosRondogiannis2019HODatalog} show that, for $k\geq1$, order-$(k+1)$ Datalog captures $k$-$\EXPTIME$ over ordered databases.
Their higher-order objects are relations; in our generators, they are operations on formulae. We adapt their higher-order counter construction in our lower bounds.
With negation and well-founded semantics, the same time classes can be captured without an input order~\cite{CharalambidisKostopoulosNomikosRondogiannis2025NegationHODatalog}.
Our finite-variable restriction prevents such an extension: even cardinality over equality alone is undefinable at every order (\cref{prop:logic-parity}).
\end{myparagraph}

\begin{myparagraph}{Higher-order queries}
Benedikt et al.~\cite{BenediktPuppisVu2015Queries} study relational algebra enriched with higher-order query transformations.
For $k\geq1$, they show that evaluating Boolean terms is $k$-$\EXPTIME$-complete at degree $2k-1$ and $k$-$\EXPSPACE$-complete at degree $2k$.
Here degree is the maximum order of a variable in the term; evaluation at unbounded degree is nonelementary.
This setting also has a direct SQL connection: the HOMES system evaluates higher-order transformations by generating SQL queries for a relational database engine~\cite{VuBenedikt2011HOMES}.
The evaluation bounds concern query and combined complexity, while each fixed query has polynomial data complexity~\cite[Section~5]{BenediktPuppisVu2015Queries}.
Our descriptive characterizations concern the properties expressible by a fixed recursive generator as the input structure varies.
\end{myparagraph}

\begin{myparagraph}{Finite-variable logics}
Our presentations belong to finite-variable infinitary logic, whose expressive power and counting limitations have been studied extensively~\cite{KolaitisVardi1992Infinitary}.
We restrict this logic to formulae described by finite higher-order presentations, obtaining an effective syntax.
Krogmeier and Madhusudan~\cite{KrogmeierMadhusudan2022Learning} study learning formulae in finite-variable logics subject to regular grammar constraints.
Their algorithms construct tree automata recognizing formulae with prescribed truth values on given finite structures.
We use the same finite-model evaluation principle and combine it with higher-order generation in our model-checking algorithms.
\end{myparagraph}

\begin{myparagraph}{Recursion schemes}
We build our decision procedures on algorithms for model checking higher-order recursion schemes~\cite{Ong2006ModelChecking,Parys2021ModelChecking}.
Kobayashi et al.~\cite{KobayashiLozesBruse2017} establish reductions between recursion-scheme model checking and higher-order fixpoint-logic model checking.
There is also a proof-theoretic use of higher-order generation: Afshari et al.~\cite{AfshariEnqvistLeigh2025Herbrand} associate schemes with pattern matching to first-order proofs, representing the computation of Herbrand expansions through cut elimination.
In our setting, a scheme presents a family of sentences whose disjunction or conjunction defines a class of structures, and we study the decision problems and expressive power of these presentations.
\end{myparagraph}


\section{Conclusions and future work}\label{sec:conclusions}

We introduced a logic of first-order theories generated by higher-order recursion schemes and studied its expressive power, model checking, and satisfiability.
For disjunctive presentations over finite ordered structures, we showed that the order of generation determines the level of the exponential-time hierarchy captured by the logic, with unrestricted first-order quantification contributing one further exponential over the existential fragment.
Using finite generators of infinitary first-order formulae, we obtained descriptive characterizations usually established through higher-order quantification and fixed points.
For satisfiability, we extended the bounded-model and coherence methods for regular theories and proved undecidability even for quantifier-free sentences.
We recovered a decidable conjunctive fragment by separating higher-order Boolean structure from independently generated regular atoms.

A natural next step is to refine the satisfiability classification between unrestricted generation and regular atomic components.
In particular, it would be useful to identify which forms of shared term generation can be admitted while retaining decidability.
A second direction is to characterize expressive power over unordered finite structures.
The counting obstruction in \cref{prop:logic-parity} prevents extending our capturing results unchanged to this setting.
For bisimulation-invariant queries, we can investigate whether the inclusions into polyadic higher-order fixpoint logic in \cref{cor:phfl-comparison} are equalities.


\section{Use of Artificial Intelligence (AI)}\label{sec:ai-use}

We used GPT-5.6 Sol and GPT-6 Astra to assist in developing the technical content of \cref{sec:sat-qf-boundary,sec:sat-regular-components}, including the definitions, theorem statements, and proofs, with the full proofs given in \cref{app:sat-qf-undecidable,app:sat-regular-atoms}.
All other technical content, including definitions, theorem statements, and proofs, was provided by the authors.
We also used these tools throughout the manuscript to assist with drafting and revising the text and with checking mathematical arguments, notation, and consistency.
The authors take full responsibility for all content and claims in this paper.

\bibliography{references}

\appendix

\section{Proofs for the logic}\label[appendix]{app:logic}

\subsection{Boolean duality}\label[appendix]{app:logic-duality}

\booleanDuality*

\begin{proof}
  For unrestricted $\FO$, we add a fresh ground start symbol with rule $S^\neg\to\neg S$.
  This generates $\{\neg\varphi:\varphi\in\cL(G)\}$ without changing the order.

  For NNF fragments, we use a two-state tree transformation: $I$ copies its input, while $D$ exchanges $\top$ with $\bot$, $\land$ with $\lor$, and $\exists x$ with $\forall x$, processing formula children in state $D$.
  On an atom, $D$ adds a negation and copies its term children in state $I$; on $\neg A$, it returns $I(A)$.
  Thus $D(\varphi)$ is the NNF of $\neg\varphi$ for every NNF formula $\varphi$.

  For order $0$, make a copy of each nonterminal for each state.
  For $k\geq1$, represent each parameter and nonterminal by its two state components, passing both components at every application.
  Define the component types by $\widehat{o}=o$ and $\widehat{\tau\to\rho}=\widehat\tau\to\widehat\tau\to\widehat\rho$.
  This translation preserves type order.
  Write $M^q:\widehat\tau$ for component $q\in\{I,D\}$ of an expression $M:\tau$; parameters and nonterminals are replaced by their corresponding copies, and $(M\,N)^q=M^q\,N^I\,N^D$.
  For each rule $F\,u_1\cdots u_r\to M$ and each $q$, we add $F^q\,u_1^I\,u_1^D\cdots u_r^I\,u_r^D\to M^q$.
  Terminal transformations are implemented by auxiliary rules of order at most $1$.
  For example, the two conjunction components have rules $C_\land^I\,u^I\,u^D\,v^I\,v^D\to\land\,u^I\,v^I$ and $C_\land^D\,u^I\,u^D\,v^I\,v^D\to\lor\,u^D\,v^D$.
  The rules for the other terminals implement the clauses above.
  The unused $D$ components of term constructors may copy their inputs in state $I$.
  Each clause uses every input child exactly once. By induction on finite derivations, we show that the resulting language is exactly $\{D(\varphi):\varphi\in\cL(G)\}$.
  To avoid repeated expansion of the application clause, introduce one auxiliary nonterminal for each subexpression of an original rule and each component.
  Its parameters are the state components of the subexpression's free parameters.
  Each use of a translated subexpression then calls its auxiliary nonterminal instead of expanding the translation again.
  These parameter types have order at most $k-1$, so the auxiliary nonterminals have order at most $k$ after eta-expansion.
  The transformed types have size $O(2^k\size{\tau})$, giving polynomial size and construction time for fixed $k$.
  We obtain the semantic equivalences from De Morgan's laws, including when $\cL(G)$ is empty.
\end{proof}

\subsection{Expressive boundaries on infinite models}\label[appendix]{app:logic-infinite-models}

On infinite models, a further restriction comes from taking conjunctions or disjunctions of first-order sentences.
Elementarily equivalent models satisfy the same first-order sentences, so they agree on every generated presentation.
We illustrate this restriction with the following non-categoricity argument~\cite[Theorem~2.3.1]{Leivant1994Higher}.

\begin{proposition}[The natural-number successor structure]\label{prop:logic-naturals}
  Over the signature consisting of a constant $c$ and a unary function $f$, no presentation in $\bigodot_{k-\HORS}\FO$ characterizes the isomorphism class of
  $\cM_{\bN}=\tuple{\bN,0,S}$, where $S(n)=n+1$.
  This holds for every $k\in\bN$ and either reading, even when restricted to countably infinite models.
\end{proposition}

\begin{proof}
  Add a fresh constant $d$ to the first-order theory of $\cM_{\bN}$, together with the sentences $d\neq f^n(c)$ for every $n\geq0$.
  Every finite subset is satisfiable in an expansion of $\cM_{\bN}$ obtained by interpreting $d$ as an element different from the finitely many excluded numerals.
  Using compactness and downward L\"owenheim--Skolem, we obtain a countably infinite model of this theory.
  Its reduct $\cM'$ to the original signature is elementarily equivalent to $\cM_{\bN}$ and has an element outside the orbit of $c$.
  Thus $\cM'$ is not isomorphic to $\cM_{\bN}$ but agrees with it on every generated presentation.
\end{proof}

Write $\FO(\LFP)$ for first-order logic with least fixed points of positive formulae, and interpret $\SO$ with full second-order semantics.

\begin{corollary}\label{cor:logic-expressive-boundaries}
  For every $k\in\bN$ and $\bigodot\in\set{\bigvee,\bigwedge}$, over all models,
  \begin{align*}
    \bigodot_{k-\HORS}\FO \subsetneq L_{\omega_1\omega}^{\omega}, \qquad
    \FO(\LFP) \nsubseteq \bigodot_{k-\HORS}\FO, \qquad
    \SO \nsubseteq \bigodot_{k-\HORS}\FO.
  \end{align*}
  These relations also hold when restricted to infinite models, or to countably infinite models.
\end{corollary}

\begin{proof}
  Let $\alpha$ assert that $f$ is injective and $c$ is not in its range.
  The sentence
  \begin{align*}
    \chi_{\bN}=\alpha\land\forall x.~\bigvee_{n\geq0}x=f^n(c)
  \end{align*}
  belongs to $L_{\omega_1\omega}^2$ and characterizes $\cM_{\bN}$ via the isomorphism $n\mapsto f^n(c)$.
  By \cref{prop:logic-naturals}, this proves that the inclusion is strict.

  The orbit condition is also expressible in $\FO(\LFP)$ by
  \begin{align*}
    \forall x.~\bigl[\lfpop_{X,y}
      \bigl(y=c\lor\exists z.~(X(z)\land y=f(z))\bigr)\bigr](x),
  \end{align*}
  and in full $\SO$ by
  \begin{align*}
    \forall X.~\Bigl(
      \bigl(X(c)\land\forall y.~(X(y)\to X(f(y)))\bigr)
      \to\forall y.~X(y)\Bigr).
  \end{align*}
  Conjoining either sentence with $\alpha$ again characterizes $\cM_{\bN}$, proving both non-inclusions in all three model classes.
\end{proof}

\subsection{Second-order interpretation on infinite models}\label[appendix]{app:logic-infinite-so}

Conversely, we can encode a presentation's syntax and derivations in full $\SO$ within any infinite domain.

\begin{restatable}[Second-order interpretation on infinite models]{proposition}{infiniteSO}\label{prop:logic-infinite-so}
  Over the class of infinite models, for every $k\in\bN$ and $\bigodot\in\set{\bigvee,\bigwedge}$, $\bigodot_{k-\HORS}\FO\subsetneq\SO$.
  The inclusion remains strict on countably infinite models.
\end{restatable}

\begin{proof}
  Fix $G$ over $(\sigma,\cV)$, with $\cV=\{x_1,\ldots,x_m\}$, and an infinite model $\cM$ with universe $U$.
  Using standard second-order coding~\cite[Sections~2.3 and~2.6]{Leivant1994Higher}, existentially quantify a subset $N\subseteq U$, an element representing zero, and relations representing the graphs of successor, addition, and multiplication on $N$.
  The relativized second-order Peano axioms and arithmetic equations force a copy of standard arithmetic, which exists in every infinite $U$.

  Encode finite syntax trees and finite derivations by elements of $N$.
  An arithmetical formula $\Gen_G(n)$, relativized to $N$, expresses that $n$ codes a sentence with a completed derivation in $G$.
  Such certificates are effectively checkable; deciding membership in $\cL(G)$ is unnecessary.

  Next existentially quantify relations
  \begin{align*}
    \Val&\subseteq N\times U^m\times U,
    &\Sat&\subseteq N\times U^m.
  \end{align*}
  Since $\sigma$ and $\cV$ are finite, the compositional evaluation and satisfaction clauses can be expressed by finitely many first-order conditions on well-formed codes.
  For instance, with arithmetically defined codes as shorthand,
  \begin{align*}
    \Sat(\ulcorner\forall x_i.\psi\urcorner,\bar a)
    \iff\forall b\in U.~
    \Sat(\ulcorner\psi\urcorner,\bar a[i\mapsto b]).
  \end{align*}
  Atomic clauses use $\Val$ and the interpretations of $\sigma$; Boolean clauses use the usual truth tables.
  By induction on syntax, we show that these relations exist and agree with ordinary evaluation.
  An assignment is represented by an $m$-tuple of elements of $U$, so both relations have fixed finite arity even when $U$ is uncountable.

  Finally, conjoin these conditions with the appropriate one of
  \begin{align*}
    &\exists n\in N.~
      \bigl(\Gen_G(n)\land
      \forall\bar a\in U^m.~\Sat(n,\bar a)\bigr),\\
    &\forall n\in N.~
      \bigl(\Gen_G(n)\to
      \forall\bar a\in U^m.~\Sat(n,\bar a)\bigr).
  \end{align*}
  By existentially quantifying the auxiliary objects, we obtain two finite $\SO$ sentences that we can effectively construct from $G$.
  On infinite models, they are equivalent to $\bigvee G$ and $\bigwedge G$, respectively.
  Because every generated formula is a sentence, its truth does not depend on the assignment.
  The same conditions also apply when $\cL(G)$ is empty.
  We conclude strictness using \cref{cor:logic-expressive-boundaries}.
\end{proof}

\section{Proofs for model checking}\label[appendix]{app:model-checking}

We regard tree automata as nondeterministic bottom-up automata on finite ranked trees.
A transition $a(q_1,\ldots,q_r)\to q$ assigns state $q$ to an $a$-labelled node whose children have states $q_1,\ldots,q_r$; acceptance requires a final state at the root.
The size of an automaton includes both its states and its transitions.

Fix $\sigma$, $\cV$, and a finite model $\cM$ with universe $U$.
Write $n=\size{U}$, $m=\size{\cV}$, and $E=U^{\cV}$ for the set of assignments, so $\size{E}=n^m$.
We first describe a syntactic check shared by both constructions.
A deterministic tree automaton records whether a subtree is a term, an atomic formula, another formula, or ill formed, together with its set of free variable names.
A variable $x$ contributes the set $\{x\}$, whereas constants and nullary formulae contribute the empty set.
At other constructors it checks the sorts of the children and takes the union of their free-variable sets; a quantifier binding $x$ additionally removes $x$.
For $\exists\FO$, it rejects universal quantifiers and permits negation only on atomic formulae.
The automaton accepts a tree only if it represents a formula with no free variables.
Since $\sigma$ and $\cV$ are fixed, this automaton has constant size.
Taking its product with either semantic automaton below therefore changes the size and construction time by only a constant factor.

\subsection{Existential first-order sentences}\label[appendix]{app:efo-theory-reg}

\existentialAutomaton*

\begin{proof}
  \begin{myparagraph}{States}
  The semantic automaton has states
  \begin{align*}
    \{T_{\nu,a}:\nu\in E,~a\in U\}
    \cup\{S_\nu,F_\nu:\nu\in E\}
    \cup\{\ast\}.
  \end{align*}
  State $T_{\nu,a}$ checks that a term evaluates to $a$ under $\nu$, state $S_\nu$ checks that a formula is true under $\nu$, and state $F_\nu$ checks falsity of an atomic formula or of $\bot$.
  Every tree has a run ending in state $\ast$, allowing us to ignore the truth value of an unselected disjunct.

  \end{myparagraph}

  \begin{myparagraph}{Transitions}
  For every $\nu\in E$, the transitions evaluating terms are
  \begin{align*}
    x &\to T_{\nu,\nu(x)},\\
    c &\to T_{\nu,c^{\cM}},\\
    f(T_{\nu,a_1},\ldots,T_{\nu,a_r})
      &\to T_{\nu,f^{\cM}(a_1,\ldots,a_r)},
  \end{align*}
  for all applicable symbols and tuples $a_1,\ldots,a_r\in U$.
  For each relation symbol $R$ and tuple $\bar a\in U^{\rk(R)}$, add
  \begin{align*}
    R(T_{\nu,a_1},\ldots,T_{\nu,a_r})
      &\to S_\nu &&\text{if }\bar a\in R^{\cM},\\
    R(T_{\nu,a_1},\ldots,T_{\nu,a_r})
      &\to F_\nu &&\text{if }\bar a\notin R^{\cM}.
  \end{align*}
  The equality constructor has transitions $=(T_{\nu,a},T_{\nu,b})\to S_\nu$ when $a=b$, and $=(T_{\nu,a},T_{\nu,b})\to F_\nu$ when $a\neq b$.

  The remaining semantic transitions are
  \begin{align*}
    \top &\to S_\nu,
    &\bot &\to F_\nu,\\
    \neg(F_\nu) &\to S_\nu,\\
    \land(S_\nu,S_\nu) &\to S_\nu,\\
    \lor(S_\nu,\ast) &\to S_\nu,\\
    \lor(\ast,S_\nu) &\to S_\nu,\\
    (\exists x)(S_{\nu[x\mapsto a]}) &\to S_\nu,
  \end{align*}
  for every $\nu\in E$, $x\in\cV$, and $a\in U$ as appropriate.
  Finally, add $b(\ast,\ldots,\ast)\to\ast$ for every terminal $b\in\Sigma_\FO$.
  There are no other transitions.

  \end{myparagraph}

  \begin{myparagraph}{Correctness and acceptance}
  We first show by induction on terms that $t$ has a run ending in $T_{\nu,a}$ if and only if its value in $\cM$ under $\nu$ is $a$.
  Using this fact in the atomic transitions, we verify the stated meanings of $S_\nu$ and $F_\nu$ on atoms.
  We then prove by induction on $\exists\FO$ formulae that
  \begin{align*}
    \varphi\text{ has a run ending in }S_\nu
      \iff \cM,\nu\vDash\varphi.
  \end{align*}
  In the conjunction case, both children must be true under the same assignment.
  In the disjunction case, one child must be true and the other always has a run ending in $\ast$.
  The existential transitions choose a witness and change only the value of the bound variable, including when that name is already bound by an enclosing quantifier.

  Choose any $\nu_0\in E$ and make $S_{\nu_0}$ the sole final state before taking the product with the syntactic automaton for $\exists\FO$ sentences.
  Such an assignment exists because $U$ is nonempty.
  Truth of a sentence is independent of the assignment, so the product recognizes exactly the required language.
  The syntactic component also checks the ignored disjuncts, excluding malformed trees and formulae with free variables.

  \end{myparagraph}

  \begin{myparagraph}{Size and construction time}
  Let $r$ be the maximum of $2$ and the arities in $\sigma$.
  There are $n^{m+1}+2n^m+1$ semantic states and $O(n^{m+r})$ transitions, with constants depending only on $\sigma$ and $\cV$.
  All transitions can be enumerated in polynomial time from the explicit tables of $\cM$.
  Since $m$ and $r$ are fixed, a polynomial $P(\size{\cM})$ bounds both the size and the construction time, as required.
  \end{myparagraph}

\end{proof}

\subsection{Unrestricted first-order sentences}\label[appendix]{app:fo-theory-reg}

\firstOrderAutomaton*

\begin{proof}
  \begin{myparagraph}{States}
  We construct a deterministic semantic automaton that records the value of a term, or the truth value of a formula, under every assignment simultaneously.
  Its states consist of term states $T_g$, indexed by functions $g:E\to U$, and formula states $B_X$, indexed by sets $X\subseteq E$.
  Their intended meanings are
  \begin{align*}
    g(\nu) &= \text{the value of the term under }\nu,\\
    X &= \setpred{\nu\in E}{\cM,\nu\vDash\varphi}.
  \end{align*}

  \end{myparagraph}

  \begin{myparagraph}{Transitions}
  At a variable $x$, the automaton returns the projection $T_{\nu\mapsto\nu(x)}$, and at a constant $c$ it returns $T_{\nu\mapsto c^{\cM}}$.
  A function symbol is evaluated pointwise:
  \begin{align*}
    f(T_{g_1},\ldots,T_{g_r}) &\to T_g,
    &g(\nu)&=f^{\cM}(g_1(\nu),\ldots,g_r(\nu)).
  \end{align*}
  A relation symbol $R$ applied to $T_{g_1},\ldots,T_{g_r}$ returns $B_X$, where
  \begin{align*}
    X=\{\nu\in E:(g_1(\nu),\ldots,g_r(\nu))\in R^{\cM}\}.
  \end{align*}
  Equality is treated in the same way, using $X=\{\nu\in E:g_1(\nu)=g_2(\nu)\}$.

  The Boolean transitions are
  \begin{align*}
    \top &\to B_E,
    &\bot &\to B_{\emptyset},\\
    \neg(B_X) &\to B_{E\setminus X},\\
    \land(B_X,B_Y) &\to B_{X\cap Y},
    &\lor(B_X,B_Y) &\to B_{X\cup Y}.
  \end{align*}
  For $x\in\cV$ and $X\subseteq E$, define
  \begin{align*}
    \exists_x X &= \{\nu\in E:\exists a\in U.~\nu[x\mapsto a]\in X\},\\
    \forall_x X &= \{\nu\in E:\forall a\in U.~\nu[x\mapsto a]\in X\}.
  \end{align*}
  The quantifier transitions are
  \begin{align*}
    (\exists x)(B_X) &\to B_{\exists_x X},
    &(\forall x)(B_X) &\to B_{\forall_x X}.
  \end{align*}
  These transitions are defined for all states of the indicated sorts; transitions with other child sorts are omitted.

  \end{myparagraph}

  \begin{myparagraph}{Correctness and acceptance}
  We prove the intended meanings of the term and formula states by structural induction.
  In particular, a sentence $\varphi$ has root state $B_E$ exactly when $\cM\vDash\varphi$.
  Make $B_E$ the sole final state and take the product with the syntactic automaton for $\FO$ sentences.
  This excludes open formulae even when they happen to be true under every assignment.

  \end{myparagraph}

  \begin{myparagraph}{Size and construction time}
  The number of semantic states is
  \begin{align*}
    n^{n^m}+2^{n^m}=2^{n^{O(1)}},
  \end{align*}
  where the constant implicit in $O(1)$ depends only on $m$.
  Each state has a table representation of polynomial length in $\size{\cM}$.
  Since the ranked alphabet and its maximum arity are fixed, there are at most $2^{\size{\cM}^{O(1)}}$ combinations of a terminal and a tuple of child states.
  Computing the output of each transition takes polynomial time by the displayed clauses, including the quantifier clauses, which examine at most $n^{m+1}$ assignment--element pairs.
  Thus a bound $2^{P(\size{\cM})}$ holds for both the automaton size and its construction time, for a polynomial $P$ depending only on $\sigma$ and $\cV$.
  \end{myparagraph}

\end{proof}

\subsection{Intersection with a finite tree automaton}\label{app:hors-automaton-nonemptiness}

\begin{proof}[Proof of Lemma~\ref{lem:hors-automaton-nonemptiness}]
  For $k=0$, we normalize $G$ to a regular tree grammar with rules $F\to a(F_1,\ldots,F_r)$ and unit rules.
  The nonterminals of the product grammar pair nonterminals of $G$ with states of $A$; its rules combine grammar rules with matching automaton transitions.
  The product grammar has size polynomial in $|G|+|A|$.
  Compute, from the bottom up, which product nonterminals can derive a finite terminal tree, and check whether one of them pairs the start nonterminal with a final state.

  For $k\geq1$, we replace the alternative right-hand sides of each nonterminal by one right-hand side joining them with a fresh binary choice terminal.
  We obtain a deterministic scheme of the same order and polynomial size, whose possibly infinite tree records all rule choices.
  We represent nonterminals without rules by divergence.
  On this tree, use an alternating parity automaton that starts by choosing a final state of $A$.
  In state $q$ at a choice node, it chooses one child and keeps $q$.
  At a node labelled $a$ of rank $r$, it chooses a transition $a(q_1,\ldots,q_r)\to q$ of $A$ and checks child $i$ in state $q_i$ for every $i$.
  A matching nullary transition accepts immediately; if there is no matching transition, the automaton rejects.
  The automaton also rejects divergence.
  Every state has an odd priority, so every infinite branch of a run is rejecting.
  These transitions are represented explicitly from those of $A$, giving a polynomial-size automaton.

  An accepting run selects one rule at each encountered choice and checks every child of each selected constructor.
  It has no infinite branch and is finitely branching, hence is finite by K\"onig's lemma.
  By removing its choice nodes, we obtain a finite tree in $\cL(G)$ together with an accepting run of $A$.
  Conversely, any accepted finite tree generated by $G$ determines such a run, so acceptance is equivalent to $\cL(G)\cap\cL(A)\neq\emptyset$.
  We now apply the order-$k$ parity model-checking bound~\cite{Ong2006ModelChecking,Parys2021ModelChecking} to obtain time $\exp_k((|G|+|A|)^{O(1)})$.
\end{proof}

\subsection{Complexity of model checking}\label{app:model-checking-complexity}

\begin{proof}[Proof of \cref{thm:model-checking-existential,thm:model-checking-first-order}]
  By \cref{prop:logic-duality} and closure of the stated complexity classes under complement, it suffices to consider the disjunctive readings.
  We construct a tree automaton $A_\cM$ recognizing exactly the sentences of the chosen fragment that are true in $\cM$.
  Then $\cM\vDash\bigvee G$ if and only if $\cL(G)\cap\cL(A_\cM)\neq\emptyset$.

  For $\exists\FO$, we use \cref{lem:efo-theory-reg} to construct an automaton of polynomial size in polynomial time in $\size{\cM}$.
  We then apply \cref{lem:hors-automaton-nonemptiness} to obtain the $k-\EXPTIME$ upper bound.
  For full $\FO$, we use \cref{lem:fo-theory-reg} to construct an automaton of size $2^{\size{\cM}^{O(1)}}$, with the same bound on its construction time.
  Constructing this automaton and testing nonemptiness together take time at most $\exp_k\bigl((\size{G}+2^{\size{\cM}^{O(1)}})^{O(1)}\bigr)\leq\exp_{k+1}\bigl((\size{G}+\size{\cM})^{O(1)}\bigr)$.
  This bound also holds when $k=0$.
\end{proof}

\section{Proofs for descriptive complexity}\label[appendix]{app:descriptive-complexity}

We first construct fixed schemes simulating machines on padded ordered word structures.
For the existential fragment, we use higher-order counters. For full first-order logic, we obtain one additional exponential by representing the base addresses as relations.
We then interpret the padded words by tuples of elements of the original input model.
In this way, we establish definability of queries on the original model, as required by the capturing statements.

\subsection{The existential lower bound}\label[appendix]{app:descriptive-counters}

We first simulate a fixed time-bounded machine on padded ordered word models.
The use of complemented higher-order counters is related to the construction for higher-order Datalog in~\cite{CharalambidisNomikosRondogiannis2019HODatalog}.
We give the rule families explicitly, including their interpretation under finite-derivation semantics.

\begin{myparagraph}{Rule schemata and open terms}
For a HORS $G$ and a closed applicative term $M:o$, write $\cM,\nu\vDash_G M$ if $M\Rightarrow_G^*\varphi$ for some finite formula tree $\varphi$ with $\cM,\nu\vDash\varphi$.
Here closedness excludes free HORS parameters, but the generated formula may have free first-order names interpreted by $\nu$.
We suppress $\nu$ when the truth value is independent of its choice.

The following schemata abbreviate ordinary curried HORS rules.
A paired argument $J=(J^0,J^1)$ abbreviates two arguments, and $F(J)$ abbreviates $F\,J^0\,J^1$.
A superscript $K$ on a scan denotes a separate family specialized to the displayed, fixed body $K$; the body is never passed as a functional argument.
Its free HORS parameters become ordinary parameters of the specialized family and are carried unchanged through recursive calls.
Before this conversion, those parameters are renamed apart from the scan's cursor parameters.
Specialization is performed once for each static body occurrence, before actual arguments are substituted; substituted expressions reuse the existing families.
The notation $\lambda L.K(L)$ below only identifies the hole in such a body.
An equation defining a formula-valued expression abbreviates a rule for its fully applied helper, and a displayed atomic test abbreviates its usual finite first-order syntax tree.

For every nonterminal $F$ of this construction with parameters $u_1,\ldots,u_r$, we also include the rule $F\,u_1\cdots u_r\to\bot$.
Consequently every closed ground expression used below has a finite unfolding, including expressions whose intended truth value is false.
No test can succeed through these alternatives, since all calls occur positively and negation is used only on input atoms.
They ensure that, for the expressions below, $\vDash_G$ commutes with conjunction and disjunction: a successful disjunct can be combined with any finite unfolding of the other disjunct.
Similarly, $\cM,\nu\vDash_G\exists x.M$ holds exactly when $\cM,\nu[x\mapsto d]\vDash_G M$ for some $d$ in the domain.

\end{myparagraph}

\begin{myparagraph}{Word models and counter types}
Let $w=a_0\cdots a_{\ell-1}$ be a word of length $\ell\geq2$, and let $N\geq\ell$.
The model $\cM_w$ has domain $D=\{0,\ldots,N-1\}$, the usual order and successor relations, endpoint constants $\ordmin,\ordmax$, the predicate $\InputLast$ selecting $\ell-1$, and a predicate $\Char_a$ for each input letter $a$.
As in the main text, $\First(x)$ and $\Last(x)$ abbreviate equality with the corresponding endpoint constant.
The letter predicates describe $w$ on the initial interval of length $\ell$ and hold nowhere else.
Put
\begin{align*}
  I_0&=D, & I_{i+1}&=\set{0,1}^{I_i}, & \size{I_i}&=\exp_i(N),\\
  \tau_0&=o, & \tau_{i+1}&=\tau_i\to\tau_i\to o, & \ord(\tau_i)&=i.
\end{align*}
We read elements of $I_{i+1}$ as binary integers, with the order on $I_i$ running from least to most significant bit.
Thus $I_i$ is identified with $\{0,\ldots,\exp_i(N)-1\}$.

For $i\geq1$, an index $J\in I_i$ is represented by two terms $J^0,J^1:\tau_{i-1}$.
Its component $J^b(L)$ tests whether its bit at $L\in I_{i-1}$ is $b$.
For $i>1$, this is the ordinary application $J^b\,L^0\,L^1$.
For $i=1$, the component is a ground expression evaluated with a distinguished address name $a$ set to $L$.
A representative is well formed when these two tests are complementary at every address.
Their truth values must be independent of scratch names; at level one the queried address is passed only through the designated address name.
Other free first-order names, when explicitly specified, are external parameters and are not rebound by counter operations.

We call the representatives obtained from the constant endpoints by predecessor at nonzero values, successor at nonmaximum values, and either saturated operation below \emph{canonical}.
In successful guarded calls, all addresses passed to bit tests and all scan boundaries are canonical.
In particular, they do not depend on the separate names used later to encode input positions.
Well-formedness and the operation invariants below concern these admissible arguments; arbitrary open terms with additional dependencies on scratch names are not admitted.

\end{myparagraph}

\begin{myparagraph}{Finite scans}
For $i\geq1$, the strict scans for a fixed body $K$ have the rules
\begin{align*}
  \AllBelow_i^K(J)
    &\to\Zero_i(J),\\
    &\to\Nonzero_i(J)\land K(\Pred_i(J))
      \land\AllBelow_i^K(\Pred_i(J)),\\
  \SomeBelow_i^K(J)
    &\to\Nonzero_i(J)\land K(\Pred_i(J)),\\
    &\to\Nonzero_i(J)
      \land\SomeBelow_i^K(\Pred_i(J)),\\
  \AllAbove_i^K(J)
    &\to\Max_i(J),\\
    &\to\Nonmax_i(J)\land K(\Succ_i(J))
      \land\AllAbove_i^K(\Succ_i(J)).
\end{align*}
The endpoint representatives are denoted $\mathbf0_i$ and $\mathbf1_i$.
Complete scans are obtained by
\begin{align*}
  \All_i^K
    &\to K(\mathbf1_i)\land\AllBelow_i^K(\mathbf1_i),\\
  \Some_i^K
    &\to K(\mathbf1_i),\\
    &\to\SomeBelow_i^K(\mathbf1_i).
\end{align*}
The complete wrappers have no boundary argument.

At level zero, a body is evaluated through the address interface $a$.
For a different name $z$, the abbreviation
\begin{align*}
  \At_z(H)=\exists a.(a=z\land H)
\end{align*}
sets that interface without renaming a substituted HORS parameter.
This is an inline formula context, not a HORS nonterminal taking $H$ as an argument.
With $z$ a scratch name distinct from all external names of $K$, the level-zero scans have the following rule schemata, with external HORS parameters suppressed:
\begin{align*}
  \AllBelow_0^K
    &\to\First(a),\\
    &\to\neg\First(a)\land
      \exists z.(\Succ(z,a)\land\At_z(K)
        \land\At_z(\AllBelow_0^K)),\\
  \SomeBelow_0^K
    &\to\neg\First(a)\land
      \exists z.(\Succ(z,a)\land\At_z(K)),\\
    &\to\neg\First(a)\land
      \exists z.(\Succ(z,a)
        \land\At_z(\SomeBelow_0^K)),\\
  \AllAbove_0^K
    &\to\Last(a),\\
    &\to\neg\Last(a)\land
      \exists z.(\Succ(a,z)\land\At_z(K)
        \land\At_z(\AllAbove_0^K)),\\
  \All_0^K
    &\to\exists a.(\Last(a)\land K
      \land\AllBelow_0^K),\\
  \Some_0^K
    &\to\exists a.(\Last(a)\land K),\\
    &\to\exists a.(\Last(a)\land\SomeBelow_0^K).
\end{align*}
These displays use no HORS parameter for a domain element or for the body.
Separate occurrences of $\At_z$ bind $a$ only in their own subtrees, so evaluating $K$ does not change the address seen by a sibling subtree.
A nested scan can reuse $z$ after the interface value has been copied into $a$.
If a fixed body needs another base value to survive, it first saves that value in a separate name outside the rebinding context.
There are only finitely many such roles, and recursive calls reuse their names after the associated guards have been emitted.

\end{myparagraph}

\begin{myparagraph}{Tests and arithmetic}
At level zero, zero, maximum, equality, and order are the input tests $\First$, $\Last$, $=$, and $<$.
Nonzero and nonmaximum are the negative endpoint tests, and predecessor and successor are expressed using $\Succ$.
Their saturated versions are the relations
\begin{align*}
  \SPred_0(x,y)
    &=(\First(x)\land x=y)\lor\Succ(y,x),\\
  \SSucc_0(x,y)
    &=(\Last(x)\land x=y)\lor\Succ(x,y).
\end{align*}
For each $i\geq0$, the endpoint components at level $i+1$ are
\begin{align*}
  \mathbf0_{i+1}^0(L)&=\top, & \mathbf0_{i+1}^1(L)&=\bot,\\
  \mathbf1_{i+1}^0(L)&=\bot, & \mathbf1_{i+1}^1(L)&=\top.
\end{align*}
For $i>0$, these are nonterminals of type $\tau_i$ ignoring their two arguments.
For well-formed $J,K\in I_{i+1}$, define
\begin{align*}
  \Same_{i+1}(J,K,L)
    &=(J^0(L)\land K^0(L))\lor(J^1(L)\land K^1(L)),\\
  \Zero_{i+1}(J)&=\All_i^{\lambda L.J^0(L)},\\
  \Max_{i+1}(J)&=\All_i^{\lambda L.J^1(L)},\\
  \Nonzero_{i+1}(J)&=\Some_i^{\lambda L.J^1(L)},\\
  \Nonmax_{i+1}(J)&=\Some_i^{\lambda L.J^0(L)},\\
  \Eq_{i+1}(J,K)
    &=\All_i^{\lambda L.\Same_{i+1}(J,K,L)},\\
  \Less_{i+1}(J,K)
    &=\Some_i^{\lambda P.J^0(P)\land K^1(P)
      \land\AllAbove_i^{\lambda L.\Same_{i+1}(J,K,L)}(P)}.
\end{align*}
The last test chooses the most significant differing position.
For $b\in\set{0,1}$, arithmetic components are given by
\begin{align*}
  (\Pred_{i+1}(J))^b(L)
    ={}&\Nonzero_{i+1}(J)\land
      \bigl((\SomeBelow_i^{\lambda Q.J^1(Q)}(L)\land J^b(L))\\
      &\hspace{28mm}\lor
        (\AllBelow_i^{\lambda Q.J^0(Q)}(L)\land J^{1-b}(L))\bigr),\\
  (\Succ_{i+1}(J))^b(L)
    ={}&\Nonmax_{i+1}(J)\land
      \bigl((\SomeBelow_i^{\lambda Q.J^0(Q)}(L)\land J^b(L))\\
      &\hspace{28mm}\lor
        (\AllBelow_i^{\lambda Q.J^1(Q)}(L)\land J^{1-b}(L))\bigr),\\
  (\SPred_{i+1}(J))^b(L)
    ={}&(\Zero_{i+1}(J)\land\mathbf0_{i+1}^b(L))
      \lor(\Pred_{i+1}(J))^b(L),\\
  (\SSucc_{i+1}(J))^b(L)
    ={}&(\Max_{i+1}(J)\land\mathbf1_{i+1}^b(L))
      \lor(\Succ_{i+1}(J))^b(L).
\end{align*}
When $i=0$, the displayed lower address is the current value of $a$, and the scans on the right are exactly the ground families above.

\end{myparagraph}

\begin{lemma}[Counter correctness and order]\label{lem:dc-counter-correctness}
  For each fixed $i$, these counter operations and scans expand to finite HORS rule families of order at most $i$, provided their external index parameters have levels at most $i$.
  On admissible well-formed representatives of $I_i$, the tests have their indicated meanings, predecessor and successor are correct away from their respective boundaries, and their saturated versions additionally fix those boundaries.
  The scans implement conjunction over, or selection from, exactly the indicated finite intervals.
  Canonical representatives are well formed and independent of external first-order names.
\end{lemma}

\begin{proof}
  We induct on the level, proving the assertions uniformly in assignments to external names.
  At level zero, the successor relation chooses the unique adjacent element, and each $\At_z$ copies that element into the address interface before continuing.
  We prove the scan assertions by induction on the interval length: the endpoint rule closes an empty universal scan, and every successful recursive rule visits the next element and strictly shortens the remaining interval.
  The existential-below scan has no successful rule at zero.

  Suppose correctness holds through level $i$.
  The endpoint pairs at level $i+1$ represent the all-zero and all-one vectors.
  The lower-level scans in the zero, maximum, and equality tests inspect exactly all their bit positions.
  The strict-order test finds a position with bits $0$ and $1$ and requires agreement at every more significant position, which is precisely binary comparison.
  In the predecessor equation, the bit at $L$ is unchanged if a less significant one-bit absorbs the borrow, and otherwise it is flipped.
  These cases are disjoint and exhaustive, and the nonzero guard excludes underflow.
  Thus the output components are complementary and represent the predecessor.
  The successor argument is the corresponding carry calculation, and the saturated equations add exactly the missing endpoint cases.
  We now prove the level-$(i+1)$ scan assertions by induction on interval length.
  Since the scans start from canonical boundaries and arithmetic preserves canonical representatives, all queried lower addresses satisfy the required invariant.

  These inductions concern successful finite unfoldings.
  In the forward direction, the guards force precisely the indicated finite traversal; in the reverse direction, unfolding that traversal and its lower-level tests produces a finite witness tree.
  We use the additional $\bot$ rules to complete unused disjuncts to finite trees; no guarded test can succeed through these rules.
  Consequently the argument applies to $\vDash_G$, without assuming that an unsuccessful recursive alternative terminates.

  For $j\geq1$, a level-$j$ component has type $\tau_{j-1}$, of order $j-1$.
  A strict scan over $I_j$ takes two such components, and hence has order $j$ before its external parameters are added.
  When we add components of levels at most $i$, the order remains at most $\max(i,j)$.
  Arithmetic components are eta-expanded with their lower-address arguments; their largest parameter order is $i-1$ at level $i$.
  At level zero the displayed families are ground when the body has no free HORS parameters; otherwise closure conversion carries those parameters, with the same order bound.
  The complete-scan wrappers carry only the same external parameters and add no boundary argument.
  There are finitely many levels and fixed body occurrences, so specialization and closure conversion produce finitely many nonterminals with the asserted order bound.
\end{proof}

\begin{myparagraph}{Embedding input positions}
The input must occupy consecutive higher-order tape positions.
Reserve names $x,y$ for the input-position encoding, disjoint from the address and scratch names of all counter scans.
At level zero, $\Code_0(x)$ is the domain value named by $x$.
For $i\geq1$, introduce component nonterminals satisfying
\begin{align*}
  \Code_i^b(x)(L)
    &\to\First(x)\land\mathbf0_i^b(L),\\
    &\to\neg\First(x)\land
      \exists y.(\Succ(y,x)\land
        \exists x.(x=y\land(\Succ_i(\Code_i(x)))^b(L))).
\end{align*}
The annotation $x$ names a first-order register and is not a HORS argument.
For $i=1$ there are no HORS arguments, while for $i>1$ the displayed $L$ abbreviates a pair of component arguments.
Every lower address passed here is canonical and therefore independent of $x,y$.
The inner binder changes only the recursive input position; it does not change the represented address $L$.
The old value of $x$ is used only in the successor guard outside that binder.

\end{myparagraph}

\begin{lemma}[Value-preserving encoding]\label{lem:dc-code-correctness}
  For every $i\geq0$ and $d\in D$, under $\nu(x)=d$ the pair $\Code_i(x)$ represents the integer $d\in I_i$ when tested at canonical lower addresses.
  The component terms and their uses in $\Eq_i(P,\Code_i(x))$ and $\Less_i(\Code_i(x),P)$, for canonical $P$, use only nonterminals of order at most $i$.
\end{lemma}

\begin{proof}
  The level-zero assertion is immediate.
  Fix $i\geq1$ and induct on $d$.
  At $d=0$, only the endpoint alternative can succeed, and it denotes the all-zero representative.
  At $d>0$, the unique witness for the outer existential quantifier is $d-1$.
  After the inner rebinding, we obtain a representative of $d-1$ at the unchanged lower address by the induction hypothesis.
  We then compute the bits of $d$ using the successor equations, because $d-1$ is below the maximum of $I_i$.
  Every successful recursive encoding step decreases $d$, so these witnesses have finite unfoldings.
  Both comparison tests enumerate their lower addresses from the constant endpoints using the counter operations, and hence satisfy the independence requirement on $x,y$.
  This also holds for every lower scan performed inside the successor equations.
  Finally, the encoding component has type $\tau_{i-1}$ and invokes only operations of order at most $i$; the first-order binders introduce no additional HORS type order.
\end{proof}

\begin{myparagraph}{Machine normalization and the initial row}
Fix a deterministic one-tape machine $\cT$, with a one-sided tape, input alphabet $A$, tape alphabet $\Gamma$, and blank symbol $\square\notin A$.
We use the usual normalization with a protected left endpoint, so the head never moves left from cell zero.
The first tape symbol carries a boundary flag, which the finite transition table preserves; this is determined by the first input position and requires no additional relation in the word model.
Write $\dot a\in\Gamma$ for the protected copy of an input symbol $a$.
Accepting and rejecting states are stationary sinks, and there is a single accepting state $q_{\mathrm{acc}}$.
Choose $N$ large enough that $\exp_k(N)-1$ bounds the running time and the head never reaches the last cell of $I_k$.
These normalizations and this padding change a time bound $\exp_k(\ell^c)$ only by increasing the fixed exponent $c$.

Let $\Gamma_{\cT}=\Gamma\cup\{(q,a):q\in Q,~a\in\Gamma\}$ be the alphabet recording a state at the head position.
The finite tape alphabet $\Gamma$ includes the protected symbols.
For each $\gamma\in\Gamma_{\cT}$, introduce a cell nonterminal with type
\begin{align*}
  \Cell_\gamma:
    \tau_{k-1}\to\tau_{k-1}\to\tau_{k-1}\to\tau_{k-1}\to o
    \qquad(k\geq1).
\end{align*}
The first pair represents the tape position $P$, and the second represents time $T$.
For $k=0$, the cell nonterminals are ground and use distinguished first-order names $p,t$.

The input tests are
\begin{align*}
  \Input_a(P)
    &=\exists x.(\Char_a(x)\land
      \Eq_k(P,\Code_k(x))),\\
  \AfterInput(P)
    &=\exists x.(\InputLast(x)\land
      \Less_k(\Code_k(x),P)).
\end{align*}
For $a\in A$, define the initial predicates as follows:
\begin{align*}
  \Init_{(q_0,\dot a)}(P)
    &=\Zero_k(P)\land
      \exists x.(\First(x)\land\Char_a(x)),\\
  \Init_a(P)&=\Nonzero_k(P)\land\Input_a(P),\\
  \Init_\square(P)&=\AfterInput(P).
\end{align*}
Every other initial predicate is $\bot$.
The initial rule is
\begin{align*}
  \Cell_\gamma(P,T)
    \to\Zero_k(T)\land\Init_\gamma(P).
\end{align*}

\end{myparagraph}

\begin{myparagraph}{Transition rules}
The machine induces a finite local update map $\delta:\Gamma_{\cT}^3\to\Gamma_{\cT}$ on valid configuration windows.
It first updates a state-marked centre cell, and otherwise checks whether a neighbouring head moves into the centre.
Its values on other windows can be chosen arbitrarily.
At an endpoint we repeat the endpoint cell as its missing neighbour.
We give priority to the centre-head case to obtain the correct update at the protected left endpoint; the right endpoint is never reached by the head.
For each triple with $\delta(\gamma_-,\gamma_0,\gamma_+)=\gamma$, add
\begin{align*}
  \Cell_\gamma(P,T)\to{}&\Nonzero_k(T)\land
    \Cell_{\gamma_-}(\SPred_k(P),\Pred_k(T))\\
    &\land\Cell_{\gamma_0}(P,\Pred_k(T))\\
    &\land\Cell_{\gamma_+}(\SSucc_k(P),\Pred_k(T)).
\end{align*}
For $k\geq1$, paired application makes this an ordinary rule with four component parameters.
Copied predecessor expressions can unfold independently, but the counter invariant assigns the same time value to every successful use.

For completeness, the order-zero instance expands as follows.
For actual names $r,s$, neither of which is $p',t'$, write
\begin{align*}
  \Call_\gamma(r,s)
    =\exists p'.\exists t'.(p'=r\land t'=s\land
      \exists p.\exists t.(p=p'\land t=t'\land\Cell_\gamma)).
\end{align*}
The saved aliases $p',t'$ ensure that even an actual argument named $p$ or $t$ retains its old value before those names are rebound.
Each displayed call is a fixed formula context, so it introduces no nonterminal with a HORS argument for a domain element.
Using separate neighbour names $l,r,u$, the transition rule is
\begin{align*}
  \Cell_\gamma\to{}&\neg\First(t)\land
    \exists l.\exists r.\exists u.
    (\SPred_0(p,l)\land\SSucc_0(p,r)\land\Succ(u,t)\\
    &\hspace{13mm}\land\Call_{\gamma_-}(l,u)
      \land\Call_{\gamma_0}(p,u)
      \land\Call_{\gamma_+}(r,u)).
\end{align*}
The initial rule is $\Cell_\gamma\to\First(t)\land\Init_\gamma(p)$.
The names $p,t,l,r,u,p',t'$ are distinct from all names reserved for counter addresses and input encoding.
Their reuse in a recursive call occurs only inside that call, after the aliases and successor guards for the enclosing call have been emitted.

\end{myparagraph}

\begin{lemma}[Cell invariant]\label{lem:dc-cell-invariant}
  For canonical $P,T\in I_k$ and $\gamma\in\Gamma_{\cT}$,
  \begin{align*}
    \cM_w\vDash_G\Cell_\gamma(P,T)
      \quad\Longleftrightarrow\quad
    \text{cell $P$ contains $\gamma$ at time $T$ in the run of $\cT$ on $w$}.
  \end{align*}
  At $k=0$, the left-hand side means $\cM_w,\nu\vDash_G\Cell_\gamma$ with $\nu(p)=P$ and $\nu(t)=T$.
\end{lemma}

\begin{proof}
  We induct on the numeric value of $T$.
  At zero, transition guards fail, and \cref{lem:dc-code-correctness} places the input symbols at exactly the consecutive positions $0,\ldots,\ell-1$.
  The initial rule marks the first symbol with the start state and boundary flag and makes every position beyond the input blank.
  Thus, at each position, the cell predicate has a successful unfolding exactly for the actual initial symbol.

  At positive time, the initial rule cannot succeed.
  By \cref{lem:dc-counter-correctness}, the three recursive premises of a transition rule refer to the correct saturated neighbours at the same time $T-1$.
  By the induction hypothesis, the three symbols form the actual window at time $T-1$; using the local update table, we then determine the current symbol.
  Conversely, we construct a finite successful unfolding by choosing that window's rule and the three finite witnesses from the induction hypothesis.
  At order zero, the successor guards and saved aliases implement precisely the same three calls.
  The additional false alternatives do not affect the argument, and every successful branch of the cell recursion decreases time until the initial row.
\end{proof}

For $k\geq1$, put $K_{\mathrm{acc}}(P)=\bigvee_{a\in\Gamma}\Cell_{(q_{\mathrm{acc}},a)}(P,\mathbf1_k)$ and use the start rule
\begin{align*}
  S\to\Some_k^{K_{\mathrm{acc}}}.
\end{align*}
At $k=0$, use the ground rule $S\to\exists p.\exists t.(\Last(t)\land\bigvee_{a\in\Gamma}\Cell_{(q_{\mathrm{acc}},a)})$.
All first-order input interfaces are bound by these rules and their called contexts, so every generated tree is an $\exists\FO$ sentence.

\begin{lemma}[Simulation on padded word models]\label{lem:descriptive-word-existential}
  Fix $k\geq0$ and a deterministic machine $\cT$ running in time $\exp_k(\ell^c)$ for some fixed $c$.
  There is a fixed $k$-$\HORS$ $G_{\cT}$ over the word vocabulary and a fixed finite name set such that, for every word of length $\ell\geq2$ and every sufficiently padded word model $\cM_w$,
  \begin{align*}
    \cM_w\vDash\bigvee G_{\cT}
      \quad\Longleftrightarrow\quad
    \cT\text{ accepts }w.
  \end{align*}
  It suffices that $\exp_k(N)-1$ bounds the normalized running time and that the normalized computation uses fewer than $\exp_k(N)-1$ tape cells; a choice $N=\ell^p$ with fixed sufficiently large $p$ satisfies both conditions.
\end{lemma}

\begin{proof}
  The start rule selects an accepting cell at the maximum time.
  By \cref{lem:dc-cell-invariant} and the stationary sink convention, such a cell exists exactly when $\cT$ accepts.
  The counter, encoding, and cell families have order at most $k$ by \cref{lem:dc-counter-correctness,lem:dc-code-correctness} and the displayed cell types.
  For $k=0$, all their parameters are carried by first-order names and the rules are ground.
  The machine, its local table, the level bound, and all specialized body occurrences are fixed, so the resulting scheme and its finite name set are independent of $w$ and $N$.
  We expand the atomic abbreviations into ordinary first-order formula trees, adding only fixed ground subtrees.
  Finally, because $\ell\geq2$, we can increase the fixed padding exponent to absorb the normalization overhead and meet the stated strict time and space margins.
\end{proof}

\subsection{Eliminating existential relation quantifiers}\label[appendix]{app:descriptive-relations}

We next give a translation that replaces existential relation quantifiers by first-order quantifiers without increasing the order of the generating scheme.
Its source language is an auxiliary language used only in the lower-bound construction.
Fix a finite set $\cV$ of first-order names, a finite set $\cQ$ of relation names of positive fixed arities, and a finite collection of literal templates.
Each template is an input literal, $P(\bar t)$, or $\neg P(\bar t)$, where $P\in\cQ$ and $\bar t$ is a fixed tuple of first-order terms over $\cV$.
The constants $\top$ and $\bot$ are also allowed as templates.
In the auxiliary ranked alphabet, each literal template is represented by a nullary terminal.
The source formulae have grammar
\begin{align*}
  \varphi
    &::= \ell
      \mid \varphi\land\varphi
      \mid \varphi\lor\varphi
      \mid \exists x.\varphi
      \mid \exists P.\varphi,
\end{align*}
where $\ell$ is a template, $x\in\cV$, and $P\in\cQ$.
Relation quantifiers have their usual second-order semantics and may be nested with first-order quantifiers.
All names use ordinary lexical binding, including when a name is rebound.
We extend the notation $\cM\vDash\bigvee G$ to such a source scheme by asking whether some generated source sentence is true in $\cM$.

The finite-template restriction lets us use one ground helper for each relation literal.
It holds for the source schemes constructed below.
After the translation, every remaining input literal terminal is replaced by its fixed formula tree over $\Sigma_\FO$.
Thus the target uses the structured first-order syntax of \cref{sec:logic}.
The translation does not require an unbounded collection of nullary terminals for arbitrary first-order terms.

Throughout this subsection, $\cM$ is a finite ordered model with domain $D$ and $\size{D}\geq2$.
We use the endpoint predicates $\First$ and $\Last$ and the successor predicate $\Succ$ of its order.
All target names introduced below are chosen outside $\cV$.
Challenge tuples, response names, and row-register tuples are pairwise disjoint, also for different source relation names.
Only finitely many new names are needed because $\cQ$ and its arities are fixed.

\begin{myparagraph}{Resolving source disjunctions}
The translation checks agreement among the literals bound by each relation quantifier.
We first arrange that every remaining literal lies in a conjunct that must be satisfied.
Define the finite set $\Prune(\varphi)$ recursively by
\begin{align*}
  \Prune(\ell)
    &= \{\ell\},\\
  \Prune(\varphi\land\psi)
    &= \{\varphi'\land\psi':
         \varphi'\in\Prune(\varphi),~
         \psi'\in\Prune(\psi)\},\\
  \Prune(\varphi\lor\psi)
    &= \Prune(\varphi)\cup\Prune(\psi),\\
  \Prune(\exists x.\varphi)
    &= \{\exists x.\varphi':\varphi'\in\Prune(\varphi)\},\\
  \Prune(\exists P.\varphi)
    &= \{\exists P.\varphi':\varphi'\in\Prune(\varphi)\}.
\end{align*}

\end{myparagraph}

\begin{lemma}\label{lem:descriptive-pruning}
  For every closed source formula $\varphi$ and every model $\cM$,
  \begin{align*}
    \cM\vDash\varphi
      \iff \cM\vDash\varphi'
        \text{ for some }\varphi'\in\Prune(\varphi).
  \end{align*}
  From an order-$k$ source scheme $G$, one can effectively construct an order-at-most-$k$ scheme $G^{\land}$ such that
  \begin{align*}
    \cL(G^{\land})
      = \bigcup_{\varphi\in\cL(G)}\Prune(\varphi).
  \end{align*}
\end{lemma}

\begin{proof}
  We prove the first claim by structural induction, with the same statement for open formulae under fixed interpretations of their free names.
  At a disjunction, select a true child; at a conjunction, retain independently chosen true prunings of both children.
  At either kind of existential quantifier, retain a witnessing element or relation and apply the induction hypothesis to its body.
  The reverse implications follow from the same clauses.
  The restriction to existential source quantifiers ensures that the choices of pruned disjuncts can be made once for the resulting finite formula.

  For the second claim, use a nondeterministic linear bottom-up tree transducer with one state $q$.
  It copies literal terminals, conjunctions, and both kinds of binders.
  At a disjunction, it has the two transitions
  \begin{align*}
    \lor((q,t_1),(q,t_2)) &\longmapsto (q,t_1),\\
    \lor((q,t_1),(q,t_2)) &\longmapsto (q,t_2),
  \end{align*}
  where $t_1,t_2$ stand for the output trees of the children.
  No output variable is copied, so this transducer is linear, although it may delete a child.
  Its images are exactly the displayed prunings.

  Finite-tree languages of higher-order recursion schemes are effectively closed under linear tree transductions without an increase in order~\cite[Theorem~2.1]{ClementeParysSalvatiWalukiewicz2016Diagonal}.
  For order $0$, the same statement is the usual closure of regular tree languages under linear tree transductions.
  We construct $G^{\land}$ using this closure property.
  The closure construction accounts for productivity of deleted subtrees, so it does not produce a pruning of an unfinished source derivation.
  This matters when one branch of a disjunction has no finite terminal unfolding.
\end{proof}

Every formula generated by $G^{\land}$ now uses only literals, conjunction, and existential quantifiers of the two sorts.
We retain ordinary first-order disjunctions inside the target gadgets defined next; these are introduced after pruning.

\begin{myparagraph}{Generating a relation table}
Fix $P\in\cQ$ of arity $s$.
Reserve challenge names $\bar i_P=(i_{P,1},\ldots,i_{P,s})$, a response name $p_P$, and two disjoint row tuples $\bar z^0,\bar z^1$ of length $s$.
The notation for the row tuples suppresses their fixed dependence on $P$.
For tuples of length $s$, equality means coordinatewise equality and inequality its negation, expanded as a disjunction of coordinate inequalities.
Define the fixed quantifier-free tests
\begin{align*}
  \First_s(\bar z)
    &= \bigwedge_{j=1}^{s}\First(z_j),\\
  \Last_s(\bar z)
    &= \bigwedge_{j=1}^{s}\Last(z_j),\\
  \Next_s(\bar z,\bar z')
    &= \bigvee_{h=1}^{s}
      \left(
        \bigwedge_{j<h}z_j=z'_j
        \land \Succ(z_h,z'_h)
        \land \bigwedge_{j>h}
          (\Last(z_j)\land\First(z'_j))
      \right).
\end{align*}
Thus $\Next_s$ is the successor relation of the lexicographic order on $D^s$, with the last coordinate least significant.
The two Boolean values are represented by the distinct endpoints:
\begin{align*}
  \beta_0(p) &= \First(p),
  &\beta_1(p) &= \Last(p).
\end{align*}

Let $b=1$ for the template $P(\bar t)$ and $b=0$ for $\neg P(\bar t)$.
For $c\in\set{0,1}$, let $\theta_{b,c}$ be $\top$ when $b=c$ and $\bot$ otherwise.
Choosing the bit $c$ for the current row $\bar z$ emits
\begin{align*}
  \Row^{b,c}_{P,\bar t}(\bar z)
    ={}& (\bar i_P\neq\bar z\lor\beta_c(p_P))\\
       &\land(\bar t\neq\bar z\lor\theta_{b,c}).
\end{align*}
The first conjunct reports the selected bit if this row is challenged.
The second requires the selected bit to equal $b$ if this row matches the value of the argument tuple $\bar t$.
All displayed tests are abbreviations for fixed first-order formula trees.

Introduce ground nonterminals $\Use^{b}_{P,\bar t}$ and $\Scan^{b,\epsilon}_{P,\bar t}$, where $\epsilon\in\set{0,1}$.
The entry rule is
\begin{align*}
  \Use^{b}_{P,\bar t}
    \to \exists\bar z^0.
      \bigl(\First_s(\bar z^0)
        \land\Scan^{b,0}_{P,\bar t}\bigr).
\end{align*}
For each $c\in\set{0,1}$ and $\epsilon\in\set{0,1}$, add a terminating rule and a continuing rule:
\begin{align*}
  \Scan^{b,\epsilon}_{P,\bar t}
    &\to \Row^{b,c}_{P,\bar t}(\bar z^\epsilon)
      \land\Last_s(\bar z^\epsilon),\\
  \Scan^{b,\epsilon}_{P,\bar t}
    &\to \Row^{b,c}_{P,\bar t}(\bar z^\epsilon)
      \land\neg\Last_s(\bar z^\epsilon)\\
    &\qquad\land\exists\bar z^{1-\epsilon}.
      \bigl(\Next_s(\bar z^\epsilon,\bar z^{1-\epsilon})
        \land\Scan^{b,1-\epsilon}_{P,\bar t}\bigr).
\end{align*}
A tuple quantifier abbreviates $s$ successive first-order quantifiers.
Negated last-tuple tests are expanded into disjunctions of negated endpoint atoms.
The row tuples alternate so that each successor test retains the current tuple while the next tuple is bound.
The indices $P,\bar t,b,\epsilon$ range over fixed finite sets and are not HORS parameters.
Consequently, every table helper has type $o$.

\end{myparagraph}

\begin{lemma}\label{lem:descriptive-relation-table}
  Let $T$ be a finite terminal expansion of $\Use^{b}_{P,\bar t}$, and let $h(T)$ be its number of rows.
  If $h(T)\neq \size{D}^s$, then $T$ is false in $\cM$ under every valuation.
  If $h(T)=\size{D}^s$, its syntactic row choices define a function $B_T:D^s\to\set{0,1}$ such that, for every valuation $\nu$ of the free first-order names,
  \begin{align*}
    \cM,\nu\vDash T
      \iff \beta_{B_T(\nu(\bar i_P))}(\nu(p_P))
        \text{ and }B_T(\bar t^{\cM,\nu})=b.
  \end{align*}
  Here $\bar t^{\cM,\nu}$ is the tuple of term values, and the endpoint assertion is evaluated in $\cM$.
  Conversely, for every $B:D^s\to\set{0,1}$, a finite expansion $T_B$ has $\size{D}^s$ rows and $B_{T_B}=B$.
\end{lemma}

\begin{proof}
  Enumerate $D^s$ in lexicographic order as $\bar d_0,\ldots,\bar d_{N-1}$, where $N=\size{D}^s$.
  The entry guard forces the first row to be $\bar d_0$.
  At each continuing rule, the successor guard forces the following row to be the unique next tuple.
  A terminating rule is possible only at $\bar d_{N-1}$, while a continuing rule at that tuple fails its negated last-tuple guard.
  Therefore the guards can all hold if and only if the expansion has exactly $N$ rows.
  This statement concerns the guards alone and does not assume that the row assertions are satisfiable.

  Suppose that $h(T)=N$, and let $c_j$ be the bit selected syntactically at its $j$th row, starting with $j=0$.
  Define $B_T(\bar d_j)=c_j$ for every $j$.
  The row-register values are uniquely fixed by the guards and are independent of $\bar i_P$, $p_P$, and the source names in $\bar t$.
  Exactly one row equals $\nu(\bar i_P)$, so the conjunction of the first row assertions is precisely $\beta_{B_T(\nu(\bar i_P))}(\nu(p_P))$.
  Exactly one row equals $\bar t^{\cM,\nu}$, so the conjunction of the second row assertions says precisely that $B_T(\bar t^{\cM,\nu})=b$.
  Conversely, if both conditions hold, use the unique row values as witnesses for the row quantifiers; every guard and every row assertion then holds.
  We have thus proved the equivalence even when no valuation satisfies its right-hand side.

  Finally, given $B$, unfold exactly $N$ rows, use a continuing rule at the first $N-1$ rows and a terminating rule at the last row, and select bit $B(\bar d_j)$ at row $j$.
  All these alternatives are available in the fixed grammar.
  The resulting expansion is $T_B$.
\end{proof}

The function $B_T$ is fixed by the chosen finite syntax and the ordered domain.
In particular, it is defined independently of the assignments later chosen for the universal challenge variables.

\begin{myparagraph}{Binding and synchronizing tables}
For a pruned source formula, define the target skeleton $\Translate(\varphi)$ by
\begin{align*}
  \Translate(\ell)
    &= \ell
      &&\text{for an input literal or }\ell\in\{\top,\bot\},\\
  \Translate(P(\bar t))
    &= \Use^{1}_{P,\bar t},\\
  \Translate(\neg P(\bar t))
    &= \Use^{0}_{P,\bar t},\\
  \Translate(\varphi\land\psi)
    &= \Translate(\varphi)\land\Translate(\psi),\\
  \Translate(\exists x.\varphi)
    &= \exists x.\Translate(\varphi),\\
  \Translate(\exists P.\varphi)
    &= \forall\bar i_P.\exists p_P.\Translate(\varphi).
\end{align*}
The last clause uses a challenge tuple of the arity of $P$.
Each occurrence of a $\Use$ helper may unfold independently.
Nested source binders for the same relation name reuse its challenge and response bank, so lexical shadowing is preserved.
The disjoint row banks ensure that the table scans capture neither these names nor any source name.

For a fixed source sentence, let $q$ range over its relation-binder occurrences and let $\Occ(q)$ be the literal occurrences bound by $q$.
Write $s_q$ for the arity of the relation bound by $q$.
When a relation $B_q$ is fixed for every $q$, write $\varphi[(B_q)_q]$ for the first-order sentence obtained by removing its relation quantifiers and evaluating each literal using the relation fixed for its binder.
Equivalently, first alpha-rename the finitely many relation-binder occurrences to distinct names and then interpret these names by the respective $B_q$.
This is notation for the proof; the generated alphabet still uses the original finite set of names.

\end{myparagraph}

\begin{lemma}\label{lem:descriptive-relation-synchronization}
  Let $\varphi$ be a closed pruned source sentence, and let $T$ be a fixed finite expansion of all table calls in $\Translate(\varphi)$.
  If a table occurrence for an $s$-ary relation has a number of rows different from $\size{D}^s$, then $\cM\not\vDash T$.
  Otherwise let $B_u$ be the function defined in \cref{lem:descriptive-relation-table} for each literal occurrence $u$.
  Then $\cM\vDash T$ if and only if all $B_u$ with $u\in\Occ(q)$ agree for each binder $q$, and $\cM\vDash\varphi[(B_q)_q]$ for the resulting common relations.
  A binder with no bound literal may be assigned any relation of its arity.
\end{lemma}

\begin{proof}
  Outside the table calls, $T$ is built using only conjunction and first-order quantifiers.
  Its domain is nonempty, so an everywhere-false table call makes the whole sentence false.
  We therefore obtain the first claim by \cref{lem:descriptive-relation-table}.

  Now assume that all table calls have the required number of rows.
  Alpha-rename the source relation-binder occurrences and the corresponding target challenge and response binders so that occurrence $q$ has names $\bar i_q,p_q$.
  The resulting source and target formulae are alpha-equivalent to the originals.
  For each $q$, define the synchronization assertion
  \begin{align*}
    C_q(\bar i_q,p_q)
      = \bigwedge_{u\in\Occ(q)}
          \beta_{B_u(\bar i_q)}(p_q),
  \end{align*}
  with an empty conjunction interpreted as $\top$.
  The displayed assertion is semantic notation on $\cM$, not an additional symbol of the target alphabet.

  Using the table lemma, we replace each expanded relation literal $u$ by the conjunction of its synchronization assertion and the query assertion that $B_u(\bar t)$ has the required polarity.
  The synchronization assertion contains no source first-order variable and no challenge or response variable of a different binder.
  The query assertion contains no challenge, response, or row variable.
  Consequently these two kinds of assertion can be separated through the surrounding conjunctions and quantifiers.
  More precisely, on a nonempty domain, if $z$ is not free in $C$, then for $Q\in\{\exists,\forall\}$,
  \begin{align*}
    Qz.(C\land\psi) \equiv C\land Qz.\psi.
  \end{align*}
  Apply this equivalence to move synchronization assertions past source binders and other challenge blocks, while keeping the block binding their own names.
  Within each such block, move the query assertions outside because they do not contain its bound names.
  Using associativity and commutativity of conjunction, we obtain the semantic factorization
  \begin{align*}
    T \equiv
      \left(\bigwedge_q\forall\bar i_q.\exists p_q.
        C_q(\bar i_q,p_q)\right)
      \land\psi_{(B_u)_u},
  \end{align*}
  where $\psi_{(B_u)_u}$ is the source first-order skeleton with each relation-literal occurrence evaluated using its own table $B_u$.

  Fix $q$ and a challenge tuple $\bar a$.
  Because the two endpoints are distinct, some response $p_q$ satisfies every conjunct of $C_q(\bar a,p_q)$ exactly when all the bits $B_u(\bar a)$ agree.
  The universal challenge ranges over all $\bar a\in D^{s_q}$, so the block for $q$ is true exactly when all tables in $\Occ(q)$ are equal pointwise.
  If $\Occ(q)$ is empty, the block is true by nonemptiness of $D$.
  Under these agreement conditions, $\psi_{(B_u)_u}$ is precisely $\varphi[(B_q)_q]$.
  Using this factorization, we conclude both directions of the claimed equivalence.
\end{proof}

\begin{proposition}\label{prop:descriptive-relation-elimination}
  Let $G$ be an order-$k$ scheme over the fixed finite source alphabet above, generating only closed source sentences.
  One can effectively construct an order-at-most-$k$ scheme $\Translate(G)$ generating first-order sentences over the same ordered signature and a fixed finite extension of $\cV$ such that, for every finite ordered model $\cM$ with at least two elements,
  \begin{align*}
    \cM\vDash\bigvee G
      \iff \cM\vDash\bigvee\Translate(G).
  \end{align*}
\end{proposition}

\begin{proof}
  We first apply \cref{lem:descriptive-pruning} to obtain $G^{\land}$.
  We then apply the fixed linear tree transduction that copies source input literals, conjunctions, and first-order binders, replaces each relation-literal terminal by its nullary $\Use$ marker, and replaces $\exists P$ by the fixed context $\forall\bar i_P.\exists p_P$.
  We obtain a scheme of order at most $k$ generating exactly the translated skeletons~\cite[Theorem~2.1]{ClementeParysSalvatiWalukiewicz2016Diagonal}.
  At order $0$ this again follows from the corresponding closure for regular tree languages.

  Replace each nullary $\Use$ marker in the scheme by its ground table nonterminal and add the rules for the table helpers.
  Since the markers are nullary and their helper rules do not call source nonterminals, this produces exactly the independent finite table expansions of their occurrences in the generated skeletons.
  Replace every remaining input-literal terminal by its fixed first-order formula tree.
  All atoms introduced by the table rules also belong to a fixed finite collection of templates, so their expansion may be postponed until after another fixed interpretation of the word vocabulary.
  These are finite ground substitutions, so they preserve the orders of existing nonterminals, and every added table nonterminal has type $o$.
  All literal occurrences use names from their source scopes, and the fresh register banks introduce no capture of source names.
  Therefore the resulting scheme $\Translate(G)$ has order at most $k$ and generates only first-order sentences.

  Suppose that $\cM\vDash\bigvee G$.
  Choose a true source sentence and a true pruning $\varphi\in\cL(G^{\land})$.
  All source quantifiers are existential, so fix a witnessing element or relation for each binder occurrence in this finite sentence.
  For each relation binder $q$, expand every table call corresponding to a literal in $\Occ(q)$ using the same witnessing relation $B_q$.
  Such expansions exist by \cref{lem:descriptive-relation-table}, and distinct calls can choose them independently.
  They determine one finite generated target sentence before its universal challenge variables are evaluated.
  The tables agree within each binder, and the chosen source witnesses satisfy $\varphi[(B_q)_q]$.
  By \cref{lem:descriptive-relation-synchronization}, we conclude that this target sentence is true.

  Conversely, suppose that some sentence $T\in\cL(\Translate(G))$ is true in $\cM$.
  Choose a translated source skeleton and table expansions producing $T$, and let $\varphi\in\cL(G^{\land})$ be its pruned source sentence.
  Every table must have the required number of rows. Using the synchronization lemma, we obtain a common relation $B_q$ for each nonvacuous binder $q$ and establish $\cM\vDash\varphi[(B_q)_q]$.
  These fixed relations witness the source relation quantifiers, while arbitrary relations witness vacuous ones, so $\cM\vDash\varphi$.
  We then recover a true sentence of $G$ using \cref{lem:descriptive-pruning}.
\end{proof}

The translation uses independent grammar choices for repeated table calls and enforces their agreement through shared first-order response variables.
Its correctness does not assume that duplicated occurrences share a nondeterministic unfolding.
The source restriction to existential first-order quantification is used both in pruning and in choosing the finite family of relation witnesses before the target challenges.

\subsection{The full first-order lower bound}\label[appendix]{app:descriptive-lift}

We obtain the additional exponential by replacing the lowest counter domain with a domain of relations.
The intermediate formulae use the existential first-order and existential relation binders of \cref{prop:descriptive-relation-elimination}.
We subsequently use that proposition to eliminate the relation binders without increasing the order of the scheme.

\begin{myparagraph}{Tuple scans and virtual values}
Let $\cM$ be an ordered word structure with domain $D$ of size $N\geq2$, and fix a positive integer $r$.
Put $U=D^r$, ordered lexicographically with the last coordinate least significant.
The predicates $\First$, $\Last$, and $\Succ$ have their usual interpretations on $D$.
Write $\FirstTuple(\bar z)$ and $\LastTuple(\bar z)$ for the conjunctions requiring every coordinate to be first and last, respectively.
The quantifier-free formula
\begin{align*}
  \LexSucc_r(\bar z,\bar z')
    =\bigvee_{h=1}^r\Bigl(
      \bigwedge_{j<h}z_j=z'_j
      \land\Succ(z_h,z'_h)
      \land\bigwedge_{j>h}
        (\Last(z_j)\land\First(z'_j))
    \Bigr)
\end{align*}
defines adjacency in $U$.
Its converse defines $\LexPred_r$.

The following displays are finite rule schemata under the specialization conventions used for the counters.
In particular, arguments denoting first-order or relation names specify one of finitely many syntactic instances; they are not additional HORS arguments.
A tuple binder abbreviates $r$ successive binders, and tuple equality abbreviates coordinatewise equality.
All temporary names are reserved separately from the free names of the arguments being inspected.

For a fixed body $K(\bar z)$, the complete tuple scan has rules
\begin{align*}
  \TupleAll^{K}
    &\to \exists\bar z^0.~
      (\FirstTuple(\bar z^0)
       \land K(\bar z^0)\land\TupleTail^{K,0}),\\
  \TupleTail^{K,\epsilon}
    &\to\LastTuple(\bar z^\epsilon),\\
  \TupleTail^{K,\epsilon}
    &\to\neg\LastTuple(\bar z^\epsilon)
       \land\exists\bar z^{1-\epsilon}.~\Bigl(
         \LexSucc_r(\bar z^\epsilon,\bar z^{1-\epsilon})\\
    &\hspace{39mm}\land K(\bar z^{1-\epsilon})
       \land\TupleTail^{K,1-\epsilon}\Bigr),
       \qquad\epsilon\in\set{0,1}.
\end{align*}
Negated endpoint tests are expanded into finite disjunctions of negated endpoint atoms, so the source remains in negation normal form.
For a given tuple $\bar p$, we define the strict-above scan by first binding $\bar z^0$ equal to $\bar p$ and then using the same tail, without testing $K$ at $\bar p$.
Thus
\begin{align*}
  \TupleAbove^{K}(\bar p)
    &\to\exists\bar z^0.~
      (\bar z^0=\bar p\land\TupleTail^{K,0}).
\end{align*}
The strict-below scan $\TupleBelow^{K}(\bar p)$ is its order dual, with $\FirstTuple$ as endpoint and $\LexPred_r$ as step relation.
Every satisfying finite expansion of these scans visits exactly the required interval: its first tuple and every subsequent tuple are uniquely determined, and its terminating rule requires the appropriate endpoint.

The virtual level-zero domain is
\begin{align*}
  V=\powerset{U}.
\end{align*}
An $r$-ary relation name $X$ represents its interpretation as a member of $V$.
We identify this interpretation with the binary integer whose bit at $\bar z$ is $X(\bar z)$, with smaller tuples less significant.
Consequently, $\size{V}=2^{N^r}$.
Satisfaction of an open generated source expression is understood relative to both a first-order assignment $\nu$ and interpretations of its free relation names.
As before, it means that some finite terminal expansion is true under those assignments.

The virtual endpoint tests are
\begin{align*}
  \VZero(X)
    &=\TupleAll^{\neg X(\bar z)},
  &\VMax(X)
    &=\TupleAll^{X(\bar z)},\\
  \VNonzero(X)
    &=\exists\bar z.~X(\bar z),
  &\VNonmax(X)
    &=\exists\bar z.~\neg X(\bar z).
\end{align*}
Here and below, a formula in a superscript specifies the fixed body of a specialized scan.
Define pointwise agreement and virtual equality by
\begin{align*}
  \VSame(X,Y,\bar z)
    &=(X(\bar z)\land Y(\bar z))
      \lor(\neg X(\bar z)\land\neg Y(\bar z)),\\
  \VEq(X,Y)
    &=\TupleAll^{\VSame(X,Y,\bar z)}.
\end{align*}
Strict order and successor have the following definitions:
\begin{align*}
  \VLess(X,Y)
    &=\exists\bar p.~\Bigl(
       \neg X(\bar p)\land Y(\bar p)
       \land\TupleAbove^{\VSame(X,Y,\bar z)}(\bar p)
      \Bigr),\\
  \VSucc(X,Y)
    &=\exists\bar p.~\Bigl(
       \neg X(\bar p)\land Y(\bar p)
       \land\TupleBelow^{X(\bar z)\land\neg Y(\bar z)}(\bar p)\\
    &\hspace{35mm}\land
       \TupleAbove^{\VSame(X,Y,\bar z)}(\bar p)
      \Bigr).
\end{align*}
Put $\VPred(X,Y)=\VSucc(Y,X)$.
The strict-order test selects the most significant differing bit.
The successor test selects the carry pivot, changes all smaller bits from $1$ to $0$, and preserves all larger bits.
Hence these tests define the order and adjacency of the integer interval $\{0,\ldots,2^{N^r}-1\}$, with no successor at its maximum and no predecessor at zero.

\end{myparagraph}

\begin{myparagraph}{Scans of the virtual domain}
Fix two alternating relation registers $X^0,X^1$, separate from the names retained by the enclosing operation.
For a fixed body $K$, define
\begin{align*}
  \VAllBelow^{K,\epsilon}
    &\to\VZero(X^\epsilon),\\
  \VAllBelow^{K,\epsilon}
    &\to\VNonzero(X^\epsilon)
       \land\exists X^{1-\epsilon}.~\Bigl(
         \VPred(X^\epsilon,X^{1-\epsilon})\\
    &\hspace{34mm}\land K(X^{1-\epsilon})
       \land\VAllBelow^{K,1-\epsilon}\Bigr),\\
  \VAllBelow^{K}(B)
    &\to\exists X^0.~
      (\VEq(X^0,B)\land\VAllBelow^{K,0}),\\
  \VAll^{K}
    &\to\exists X^0.~
      (\VMax(X^0)\land K(X^0)
       \land\VAllBelow^{K,0}).
\end{align*}
The strict-above family is obtained by exchanging $\VZero$ with $\VMax$, $\VNonzero$ with $\VNonmax$, and $\VPred$ with $\VSucc$.
Existential scans can choose a virtual value directly:
\begin{align*}
  \VSome^{K}
    &\to\exists X.~K(X),\\
  \VSomeBelow^{K}(B)
    &\to\exists X.~
      (\VLess(X,B)\land K(X)),\\
  \VSomeAbove^{K}(B)
    &\to\exists X.~
      (\VLess(B,X)\land K(X)).
\end{align*}
These are logical existential relation binders, whose interpretations range over all of $V$.
They do not bind HORS parameters.

We retain the false alternative for every formula-generating helper: a nonterminal with parameters $u_1,\ldots,u_s$ also has the rule $F\,u_1\cdots u_s\to\bot$.
This guarantees a finite, possibly false, expansion whenever an unused disjunct must be completed to a terminal formula.
All occurrences of helper computations are positive, and negations apply only to input or relation literals, so no computational test can succeed through a false alternative.

\end{myparagraph}

\begin{lemma}\label{lem:descriptive-virtual-base}
  Relative to fixed interpretations of their free names, the virtual tests define their indicated relations on $V$, and the virtual scans define existential or universal quantification over their indicated intervals.
  When the specialized body has no free HORS parameters, all these helpers have order zero.
  If a specialized body carries components of indices of levels at most $i$, its closure-converted helpers have order at most $i$.
\end{lemma}

\begin{proof}
  We prove the tuple-scan invariant by induction on the number of tuples remaining before its endpoint.
  The endpoint alternative is true exactly when no further tuple is required.
  Otherwise the step relation determines the next tuple uniquely, and the recursive alternative requires both its body and the remaining scan.
  We thus prove both soundness of every satisfying finite expansion and existence of a satisfying expansion when all required body instances hold.
  The same argument applies to the strict scans.

  We verify the endpoint and equality tests by applying this invariant to the characteristic bits.
  For order, all bits above the selected $0/1$ pivot agree, which is exactly unsigned binary comparison.
  For successor, the required lower $1/0$ bits, pivot $0/1$ bit, and unchanged higher bits are exactly binary increment without overflow.
  The positive nonendpoint tests have the claimed meanings because they choose a witnessing tuple.

  At a nonzero virtual value, the predecessor relation determines a unique next value.
  We therefore prove correctness of $\VAllBelow$ by induction on the numeric value of the current register.
  Its true endpoint alternative occurs exactly at zero, while its other true alternative decreases the value by one and checks the body there.
  The complete scan additionally checks the maximum, and the above scan is proved dually.
  The existential scans have the claimed meanings by the range of their relation quantifiers and correctness of $\VLess$.

  No common derivation is required for distinct body occurrences.
  For completeness, the scanned interval is finite, so one can choose a satisfying finite expansion for every required body instance and join these expansions into a finite derivation.
  All source quantifiers are existential; their witnesses can be chosen together with these expansions, and the guards fix every scanned value.
  For soundness, any true expansion determines exactly such witnesses and steps.

  Logical binders and fixed relation-literal templates introduce no HORS arguments.
  Thus helpers with no external HORS parameters have ground type.
  With external components, specialization threads precisely those components as ordinary parameters.
  A component of a level-$j$ index has type $\tau_{j-1}$ of order $j-1$, so we retain order at most $i$ when carrying components of levels at most $i$.
  Only finitely many specialized bodies and register roles occur, so these schemata define a finite family.
\end{proof}

\begin{myparagraph}{The lifted counter interface}
Replace the counter hierarchy by
\begin{align*}
  I_0=V,\qquad I_{i+1}=\set{0,1}^{I_i}.
\end{align*}
Then $\size{I_i}=\exp_{i+1}(N^r)$.
Keep the types $\tau_0=o$ and $\tau_{i+1}=\tau_i\to\tau_i\to o$.
For $i\geq1$, a level-$i$ index is still represented by two components of type $\tau_{i-1}$.
At level zero an index is now the interpretation of a relation register.

The only change to component application occurs at level one.
A level-one component is ground and is evaluated with a distinguished relation register $A$ denoting its queried virtual address.
To evaluate this component $J^b$ at a register $X$, the specialization emits
\begin{align*}
  J^b[X]=\exists A.~(\VEq(A,X)\land J^b),
\end{align*}
where $X$ is distinct from $A$.
If the argument value is already in $A$, we first copy it into a reserved temporary relation register before rebinding $A$.
The equality test lies outside the component it prepares, and its tuple-scan registers are separate from the component's retained registers.
At higher levels, component application remains ordinary application to the two components of the lower address.

Interpret the level-zero tests, strict scans, and complete scans by their virtual counterparts.
Use the same constants, bit comparison, and ripple-carry equations at every positive level as in \cref{lem:dc-counter-correctness}.
The saturated predecessor and successor relations at level zero are
\begin{align*}
  \VSPred(X,Y)
    &=(\VZero(X)\land\VEq(X,Y))
      \lor\VPred(X,Y),\\
  \VSSucc(X,Y)
    &=(\VMax(X)\land\VEq(X,Y))
      \lor\VSucc(X,Y).
\end{align*}
They fix the relevant endpoint and otherwise move by one.

We now repeat the counter proof with \cref{lem:descriptive-virtual-base} as its base case.
In the induction, we use only the lower domain's order, endpoints, adjacency, scans, and correct evaluation of a component at a given lower address.
We have just constructed each of these virtual operations and verified its required properties.
In particular, the constructed canonical representatives, obtained from the constant vectors by in-range predecessor and successor steps and their saturated variants, are complementary bit representations of their claimed integers.
Their only free query names are the designated address registers; they do not depend on the separate first-order registers used for input-position encoding.
This is the class of lower-address representatives used in the input-code tests below.
No additional claim about arbitrary components containing those input registers is needed.

The order calculation is also preserved.
A scan inside a level-$i$ operation carries components of levels at most $i$ and therefore has order at most $i$, including when its scanned domain is $V$.
The contexts that copy relation registers take no HORS parameters other than those same external components.
Thus they introduce no additional type order.

\end{myparagraph}

\begin{myparagraph}{Input positions and the machine simulation}
Input positions must retain their numeric values in the larger counter domain.
For $d\in D$ and a virtual register $B$, the test $\VCode(d,B)$ asserts that $B$ is the integer equal to the rank of $d$ in $D$.
Use two pairs of first-order and relation registers $(d^0,R^0)$ and $(d^1,R^1)$, separate from the address and cell registers, and define
\begin{align*}
  \VCode(d,B)
    &\to\exists d^0.\exists R^0.~
      (d^0=d\land\VEq(R^0,B)\land\VCode^{0}),\\
  \VCode^{\epsilon}
    &\to\First(d^\epsilon)\land\VZero(R^\epsilon),\\
  \VCode^{\epsilon}
    &\to\neg\First(d^\epsilon)
      \land\exists d^{1-\epsilon}.\exists R^{1-\epsilon}.~\Bigl(
        \Succ(d^{1-\epsilon},d^\epsilon)\\
    &\hspace{31mm}\land\VSucc(R^{1-\epsilon},R^\epsilon)
      \land\VCode^{1-\epsilon}\Bigr).
\end{align*}
We use the displayed entry rule when its argument names are distinct from the reserved temporary names.
For other calls, we first copy the argument values to the designated aliases outside that reserve.
The recursive alternative decreases both values by one, and the endpoint alternative identifies their zeros.
By induction on the rank of $d$, we show that the test is true exactly for the claimed virtual code.
All these codes exist because $N-1<2^{N^r}$.

For $i\geq1$, retain the value-preserving construction $\Code_i(d)$ from \cref{lem:dc-code-correctness}, using the lifted zero and successor operations.
Its zero branch emits the all-zero vector, and its recursive branch applies one lifted successor after moving the first-order input register to its unique predecessor.
The register for this recursion is separate from all scan-address registers.
Whenever we query the component, we pass a canonical lower-address representative independent of this input register.
Consequently, rebinding the input register leaves the queried address unchanged, and we prove the value-preserving property by induction on the rank of $d$ as before.
The equality and comparison tests used to initialize the tape query precisely such canonical addresses, so this restricted invariant covers every use of $\Code_i$ in the simulation.

At order zero, input-symbol and beyond-input tests can be written explicitly as
\begin{align*}
  \Input_a(P)
    &=\exists d.~(\Char_a(d)\land\VCode(d,P)),\\
  \AfterInput(P)
    &=\exists d.\exists B.~\bigl(
       \InputLast(d)\land\VCode(d,B)
       \land\VLess(B,P)\bigr).
\end{align*}
For positive orders, use the corresponding equality and comparison with $\Code_k(d)$ as in the original cell construction.
These tests place the input symbols at the consecutive positions $0,\ldots,\ell-1$ and blanks at all larger positions.

For $k\geq1$, the cell nonterminals retain their four component arguments and their order-$k$ types.
Their initial and local transition rules are those of \cref{lem:dc-cell-invariant}, with the lifted operations substituted.
For $k=0$, the current position and time are stored in relation registers $P,T$.
A transition existentially chooses relation registers $P_-,P_+,T_-$, requires
\begin{align*}
  \VNonzero(T)
  \land\VSPred(P,P_-)
  \land\VSSucc(P,P_+)
  \land\VPred(T,T_-),
\end{align*}
and conjoins the three recursive cell calls prescribed by the local transition table.
To call a cell with actual registers $R,S$, first preserve both values and then rebind the cell registers:
\begin{align*}
  \Call_\gamma(R,S)
    =\exists B_P.\exists B_T.~\Bigl(
       \VEq(B_P,R)\land\VEq(B_T,S)
       \land\exists P.\exists T.~\bigl(
          \VEq(P,B_P)\\
       \hspace{47mm}\land\VEq(T,B_T)
          \land\Cell_\gamma\bigr)\Bigr).
\end{align*}
The aliases $B_P,B_T$ are distinct from all possible actual registers.
In particular, this context also works for the middle call whose actual position is the old $P$.
Recursive uses may shadow the aliases only inside the recursive call, after the copies have been checked.
The initial rule requires $\VZero(T)$ and the appropriate initial-position test.
All these order-zero contexts use logical relation binders and ground tests.
With $\Gamma_{\mathrm{acc}}$ the finite set of cell symbols carrying the accepting state, the order-zero start rule is
\begin{align*}
  S\to\exists P.\exists T.~
    \Bigl(\VMax(T)
      \land\bigvee_{\gamma\in\Gamma_{\mathrm{acc}}}\Cell_\gamma\Bigr).
\end{align*}

We now extend the cell invariant to the lifted hierarchy for every $k\geq0$.
At time zero, we recover the initial configuration from the numeric input embedding.
At positive time, all three premises refer to the unique preceding time and to the required saturated neighboring positions.
By induction on time, we show that every successful local rule uses the actual preceding window; conversely, we construct a successful finite unfolding from that window.
The induction terminates at zero, regardless of the size of the finite counter domain.
The false alternatives cannot establish a true cell premise.

\end{myparagraph}

\begin{lemma}\label{lem:descriptive-word-first-order}
  Fix $k\geq0$ and a deterministic machine $M$ deciding a word language in time $\exp_{k+1}(\ell^c)$ for a fixed constant $c$.
  There are a constant $d\geq1$ and a fixed $k$-$\HORS$ $G_M$ generating full first-order sentences such that, for every word $w$ of length $\ell\geq2$ and every padded ordered word structure $\cM_{w,N}$ with $N\geq\ell^d$,
  \begin{align*}
    M\text{ accepts }w
      \iff\cM_{w,N}\vDash\bigvee G_M.
  \end{align*}
  Here the input occupies the first $\ell$ domain positions, $\InputLast$ identifies its last position, and the other word and order predicates have their standard meanings.
\end{lemma}

\begin{proof}
  Normalize the machine as in the cell construction, using stationary accepting and rejecting sinks and a fixed local transition table.
  Choose the padding constant $d$ large enough that $\exp_{k+1}(N^r)-1$ bounds the normalized running time and strictly exceeds every used tape position whenever $N\geq\ell^d$ and $\ell\geq2$.
  The relation arity $r$ is fixed; in particular, $r=1$ suffices when $d$ is enlarged.

  Construct the lifted source scheme over $I_0=\powerset{D^r}$ and $I_{i+1}=\set{0,1}^{I_i}$.
  Its final time is the maximum element of $I_k$, whose domain has size $\exp_{k+1}(N^r)$.
  Its start expression existentially scans the positions and asks whether an accepting cell symbol occurs at that final time.
  Using the lifted cell invariant, we show that some generated source sentence is true if and only if $M$ accepts $w$.
  Stationary sinks ensure that acceptance at this padded final time agrees with the original halting answer.

  The new ground-register operations have order zero, and a specialized helper carrying level-$i$ components has order at most $i$ by \cref{lem:descriptive-virtual-base}.
  The positive-level counter and cell types are unchanged.
  Thus the source scheme has order at most $k$, including when $k=0$.
  There are only finitely many register roles, component types, specialized bodies, and machine rules because $M$, $k$, and $r$ are fixed.
  In particular, every relation atom is one of finitely many templates over a fixed finite first-order-name set, and every relation name has the fixed positive arity $r$.

  We apply \cref{prop:descriptive-relation-elimination} to this source scheme.
  We obtain a fixed scheme $G_M$ of order at most $k$ generating full first-order sentences with the same disjunctive truth on every finite ordered structure of size at least two.
  Its generated relation tables are chosen in the syntax before their universal first-order challenges are evaluated, as required by that proposition.
  Combining the two equivalences, we conclude the lemma for every stated padding size.
\end{proof}

\subsection{Returning to the original ordered models}\label[appendix]{app:descriptive-tuples}

The preceding constructions allow a polynomially padded word domain.
We now implement that domain using a fixed number of first-order names over the original model.
In particular, the padding does not require extra elements in the input model.

\begin{myparagraph}{A word encoding by tuples}
Fix an ordered signature $\sigma$ and let $\cM$ have universe $U$ of size $n\geq2$.
There is a constant $d$ and a canonical word $w_\cM$ of length $\ell=n^d$, indexed in lexicographic order by $U^d$, such that each of its letter predicates is given by a fixed quantifier-free formula over $\cM$.
We give the encoding explicitly to ensure that this translation also preserves the existential fragment.

List the relation tables, constant graphs, and function graphs of the non-order symbols of $\sigma$ as $a_1,\ldots,a_h$.
The corresponding table predicates have the forms $R(\bar v)$, $v=c$, or $f(\bar v)=v'$, respectively.
Let $s\geq1$ bound their arities, and choose $t\geq1$ such that $2^t>h$.
Assign distinct nonzero patterns $\eta_j\in\set{0,1}^t$ to the tables, and set $d=t+s$.
For a tuple $\bar u\in U^t$, let $\Tag_j(\bar u)$ require $u_i=\ordmin$ when $\eta_j(i)=0$ and $u_i=\ordmax$ when $\eta_j(i)=1$.
These tags are disjoint because the two endpoints are distinct.
If table $a_j$ has arity $s_j$, define $\Data_j(\bar v)$ to be its table predicate on the first $s_j$ coordinates, conjoined with $v_i=\ordmin$ for all remaining coordinates.

Use the input alphabet $\{\#,0,1\}$ and define its letter predicates on $\bar z=(\bar u,\bar v)\in U^{t+s}$ by
\begin{align*}
  \lambda_\#(\bar z)
    &= \bigwedge_{i=1}^{d}z_i=\ordmin,\\
  \lambda_1(\bar u,\bar v)
    &= \bigvee_{j=1}^{h}
       \bigl(\Tag_j(\bar u)\land\Data_j(\bar v)\bigr),\\
  \lambda_0(\bar z)
    &= \neg\bigl(\lambda_\#(\bar z)\lor\lambda_1(\bar z)\bigr).
\end{align*}
All these formulae are fixed and quantifier free; we put the last one in NNF.
No nonzero tag matches the first tuple, so the word starts with a unique protected marker $\#$.
Its remaining positions record the table bits, with $0$ at unused positions.
We recover the ordered model in polynomial time by computing $n$ from $\ell=n^d$, reading its tables from their tagged positions, and reconstructing the uniquely determined order symbols.
The encoding and its inverse on valid encodings have polynomial cost for the fixed signature.

\end{myparagraph}

\begin{myparagraph}{Padding within a tuple domain}
Choose a fixed integer $q\geq d$ and interpret the padded word structure $\mathcal I_q(\cM)$ on $D=U^q$.
Its order is lexicographic, with endpoint predicates
\begin{align*}
  \First_q(\bar x)&=\bigwedge_{i=1}^{q}x_i=\ordmin,
  &\Last_q(\bar x)&=\bigwedge_{i=1}^{q}x_i=\ordmax.
\end{align*}
The strict order and successor on tuples are defined by
\begin{align*}
  \Less_q(\bar x,\bar y)
    &=\bigvee_{j=1}^{q}
      \left(\bigwedge_{i<j}x_i=y_i\land x_j<y_j\right),\\
  \Next_q(\bar x,\bar y)
    &=\bigvee_{j=1}^{q}
      \left(\bigwedge_{i<j}x_i=y_i\land\Succ(x_j,y_j)
        \land\bigwedge_{i>j}(x_i=\ordmax\land y_i=\ordmin)\right).
\end{align*}
All four predicates, and their negations in NNF, are quantifier free.

The input occupies the initial segment whose first $q-d$ coordinates are $\ordmin$.
For each $a\in\{\#,0,1\}$, put
\begin{align*}
  \Char_a(\bar z)
    &=\bigwedge_{i=1}^{q-d}z_i=\ordmin
      \land\lambda_a(z_{q-d+1},\ldots,z_q),\\
  \InputLast(\bar z)
    &=\bigwedge_{i=1}^{q-d}z_i=\ordmin
      \land\bigwedge_{i=q-d+1}^{q}z_i=\ordmax.
\end{align*}
Thus the input word is exactly $w_\cM$, its last position has numeric value $\ell-1$, and the available domain has size $N=n^q$.
We obtain any required fixed polynomial padding in $\ell$ by increasing $q$; we may take $q$ to be a multiple of $d$ when an exact power of $\ell$ is convenient.

\end{myparagraph}

\begin{lemma}[Tuple translation]\label{lem:descriptive-tuple-translation}
  Fix $q$ and a $k-\HORS$ $G$ over the padded word vocabulary, using a fixed finite set of atomic templates.
  There is a $k-\HORS$ $G^{\mathcal I_q}$ over $\sigma$ such that, for every ordered $\cM$ with at least two elements,
  \begin{align*}
    \cM\vDash\bigvee G^{\mathcal I_q}
      \iff \mathcal I_q(\cM)\vDash\bigvee G.
  \end{align*}
  The translation preserves both the $\exists\FO$ fragment and unrestricted $\FO$.
\end{lemma}

\begin{proof}
  Replace each first-order name $x$ by a distinct tuple of names $\bar x=(x_1,\ldots,x_q)$.
  Translate an atomic word predicate by its defining formula above, and translate equality by componentwise equality.
  Endpoint constants in a template are represented by tuples whose coordinates are all the corresponding endpoint of $U$.
  Translate negative atoms by the NNF negations of these quantifier-free formulae.
  Boolean connectives are preserved, and quantifiers are expanded as
  \begin{align*}
    (\exists x.~\varphi)^{\mathcal I_q}
      &=\exists x_1\cdots\exists x_q.~\varphi^{\mathcal I_q},\\
    (\forall x.~\varphi)^{\mathcal I_q}
      &=\forall x_1\cdots\forall x_q.~\varphi^{\mathcal I_q}.
  \end{align*}
  All tuples of $U^q$ are in the interpreted domain, so no relativization is needed.
  We prove preservation of truth under the corresponding assignments by structural induction.

  In the preceding constructions, postpone expansion of the finitely many atomic templates until this tuple translation is complete.
  The translation is then a fixed tree homomorphism: a literal leaf is replaced by a fixed formula tree, and a quantifier node by a chain of $q$ binders containing its translated child exactly once.
  For $k\geq1$, implement each constructor context by an auxiliary nonterminal of order at most $1$, and replace terminal occurrences by these helpers throughout the scheme.
  For $k=0$, substitute the contexts directly in the rules of the regular tree grammar.
  In both cases, we preserve order and generate exactly the translated finite tree language.
  The name tuples preserve lexical scope and shadowing, so sentences remain sentences.
  Only a fixed finite number of names is introduced, and no universal quantifier is introduced when the source is existential.
\end{proof}

\begin{proposition}\label{prop:descriptive-lower-bounds}
  Fix an ordered signature $\sigma$ and $k\in\bN$.
  Every query in $k-\EXPTIME$ over finite ordered $\sigma$-models is definable in $\bigvee_{k-\HORS}\exists\FO$.
  Every query in $(k+1)-\EXPTIME$ over these models is definable in $\bigvee_{k-\HORS}\FO$.
\end{proposition}

\begin{proof}
  Let $Q$ be a query in $k-\EXPTIME$.
  The language of words $w_\cM$ encoding its models of size at least two is decidable in $k-\EXPTIME$, since validity of this encoding and conversion to the canonical input tables take polynomial time.
  We fix a machine deciding that word language and choose $q$ large enough that the domain of $\mathcal I_q(\cM)$ meets the padding requirement of \cref{lem:descriptive-word-existential}.
  Using that lemma and \cref{lem:descriptive-tuple-translation}, we construct a fixed $k-\HORS$ $G_Q$ over $\sigma$ such that
  \begin{align*}
    \cM\in Q
      \iff \cM\vDash\bigvee G_Q
  \end{align*}
  whenever $\size{U}\geq2$.

  There are only finitely many one-element $\sigma$-models up to isomorphism.
  For each one belonging to $Q$, its complete atomic diagram is a quantifier-free sentence: use $\ordmin=\ordmax$ and the appropriate positive or negative instance of each relation symbol on $\ordmin$.
  Constants and functions have their unique possible values on a singleton.
  Add these finitely many diagram sentences as ground start alternatives, and guard the original start alternative by $\ordmin\neq\ordmax$.
  We thus obtain the required presentation on all nonempty ordered models, preserving the order and fragment.

  For a query in $(k+1)-\EXPTIME$, use \cref{lem:descriptive-word-first-order} in place of \cref{lem:descriptive-word-existential}, then apply the same tuple translation and singleton treatment.
  In both cases the resulting scheme, signature, and finite name set are fixed by $Q$ and do not depend on the model.
\end{proof}

For the full first-order result, the extra order symbols may also be eliminated.
Existentially bind two fresh first-order names $u,v$ at the start, require $\forall z.~\neg(z<u)$ and $\forall z.~\neg(v<z)$, and replace $\ordmin,\ordmax$ by $u,v$ throughout the fixed atomic templates.
Replace each successor test by $x<y\land\forall z.~\neg(x<z\land z<y)$, choosing its bound name fresh for that template.
These fixed formula contexts and the new ground start rule preserve the scheme order.
Thus the full first-order characterization also holds when the only distinguished order symbol is $<$.

\subsection{Comparison with higher-order least fixed-point logics}\label[appendix]{app:descriptive-fixed-point}\label[appendix]{sec:comparison-fixed-point}

Write $\FO(\LFP)$ for first-order logic with least fixed points of positive formulae, and $\HO^r(\LFP)$ for order-$r$ logic with least fixed points, with $\HO^1(\LFP)=\FO(\LFP)$.
Over finite ordered models, $\HO^{k+1}(\LFP)$ captures $k$-$\EXPTIME$~\cite{Immerman1999Descriptive,FreireMartins2011,BruseKronenbergerLange2022}. Combining this with \cref{thm:descriptive-existential,thm:descriptive-first-order}, we obtain the following comparison.

\begin{corollary}\label{cor:logic-comparison}
  Over finite ordered models, for every $k\in\bN$,
  \begin{align*}
    \bigvee_{k-\HORS}\exists\FO \ordequiv \HO^{k+1}(\LFP),\qquad
    \bigvee_{k-\HORS}\FO \ordequiv \HO^{k+2}(\LFP).
  \end{align*}
  Here $\ordequiv$ means equality of the Boolean queries defined on the original finite ordered models; the finite variable-name set may depend on the query.
\end{corollary}

By \cref{prop:logic-duality} and closure under complement, the first comparison also applies to $\bigwedge_{k-\HORS}\forall\FO$ and the second to $\bigwedge_{k-\HORS}\FO$.
These comparisons do not assert translations preserving syntax, variable bounds, or presentation size.
Also, every fixed $\SO$ sentence, under full second-order semantics, is equivalent on these models to a presentation in $\bigvee_{0-\HORS}\FO$: its evaluation uses polynomial space~\cite{Immerman1999Descriptive}, hence exponential time.

\subsection{Comparison with polyadic higher-order fixpoint logic}\label[appendix]{app:descriptive-phfl}
Higher-order fixpoint logic ($\HFL$) extends the modal $\mu$-calculus by higher-order functions on predicates.
Its polyadic extension $\PHFL$ interprets formulas on tuples of states of a labelled transition system.
Over finite systems, $\PHFL^{k}$ captures the bisimulation-invariant queries in $k$-$\EXPTIME$ for every $k\geq0$~\cite[Theorem~15]{BruseKronenbergerLange2022}.
Here the type order is bounded by $k$, while the tuple width may be any fixed finite number.

For comparison with our logics, represent each action by a binary relation, each proposition by a unary relation, and the distinguished state by a constant.
Write $\Inv_{\sim}(\bL)$ for the queries definable in $\bL$ that are invariant under bisimulations relating the distinguished states of two finite pointed systems.
\begin{corollary}\label{cor:phfl-comparison}
  Over finite pointed systems, for every $k\in\bN$,
  \begin{align*}
    \Inv_{\sim}\bigl(\bigvee_{k-\HORS}\exists\FO\bigr) \subseteq \PHFL^{k},\qquad
    \Inv_{\sim}\bigl(\bigvee_{k-\HORS}\FO\bigr) \subseteq \PHFL^{k+1}.
  \end{align*}
  These inclusions concern queries on the same systems without an input order; target formulas may use additional coordinates initialised to the distinguished state.
  No converse inclusion is asserted.
\end{corollary}

\begin{proof}
  With each presentation fixed, \cref{thm:model-checking-existential,thm:model-checking-first-order} put the queries on the left in $k$-$\EXPTIME$ and $(k+1)$-$\EXPTIME$, respectively.
  We then derive the inclusions using the $\PHFL$ capturing theorem and bisimulation invariance.
\end{proof}

\section{Proofs for satisfiability}\label[appendix]{app:satisfiability}

\subsection{Inherited undecidability}\label[appendix]{app:sat-inherited-undecidability}

For every $k\in\bN$, satisfiability under either reading is undecidable for $k$-$\HORS$ generating sentences in either of the following two fragments.
Both fragments allow equality and finitely many constants.
\begin{enumerate}
  \item Sentences $\varphi_\exists\land\varphi_\forall$, where $\varphi_\exists\in\exists\FO$ and $\varphi_\forall\in\forall\FO$ has at most three quantifier occurrences, over signatures consisting of constants and at most two binary relation symbols.
  \item $\exists\FO$ sentences over signatures consisting of constants and at most two unary function symbols.
\end{enumerate}
These are the undecidability results of Mathur et al.~\cite[Theorems~1.2 and~1.3]{Mathur2025Theories}.
Replacing free source parameters by fresh constants preserves satisfiability under both readings.
The results hold at every order because every regular tree grammar is a $k$-$\HORS$.

\subsection{Bounded model properties}\label[appendix]{app:sat-bounded-models}

\boundedModels*

\begin{proof}
  Fix an input scheme $G$ over $(\sigma,\cV)$ and put $T=\cL(G)$.
  We adapt the argument of~\cite[Theorem~1.1]{Mathur2025Theories}, using our finite-model checking procedure and applying the bound to $\bL^\land$ in the conjunctive case.

  Suppose first that $\bL^\land$ has the weak bounded model property, with bound $b=b(\sigma,\cV)$.
  If $T$ has a model, then every finite subset of $T$ has a model of size at most $b$, because its conjunction belongs to $\bL^\land$ and uses the same signature and variable names.
  There are only finitely many $\sigma$-models of size at most $b$ up to isomorphism.
  If none satisfied all of $T$, choose for each such model $\cN$ a sentence $\varphi_{\cN}\in T$ that it falsifies.
  The finite conjunction of these sentences would be satisfiable but have no model of size at most $b$, a contradiction.
  Thus $T$ has a common finite model whenever it has any model.

  To decide this without computing $b$, we dovetail two searches.
  The first enumerates finite $\sigma$-models and checks $\cM\vDash\bigwedge G$ using \cref{thm:model-checking-first-order}.
  The second enumerates finite derivations of $G$ and first-order refutations of finite sets of their completed outputs.
  Such derivations and refutations are effectively checkable.
  The first search terminates if $T$ is satisfiable, by the preceding argument.
  If $T$ is unsatisfiable, we obtain an unsatisfiable finite subset by compactness and a finite refutation by completeness, so the second search terminates.
  The second search enumerates finite certificates directly, so it never has to determine whether another generated sentence exists.
  This also handles finite and empty generated languages.

  Finally, suppose that $\bL$ has the strong bounded model property.
  Compute $b(\sigma,\cV)$ and enumerate all $\sigma$-models of size at most this bound.
  Some generated sentence is satisfiable if and only if $\cM\vDash\bigvee G$ for one of these models, which is decidable by \cref{thm:model-checking-first-order}.
  This finite search needs no closure of $\bL$ under conjunction.
\end{proof}

\subsection{Coherent formulae}\label[appendix]{app:sat-coherence}

We use the value-tracking formulation underlying the proof of
\cite[Proposition~7.8 and Lemma~7.9]{Mathur2025Theories}.
We define a local transport relation below and represent tracked values by the equivalence classes it generates.
Fix an algebraic signature $\sigma=\tuple{\cC,\emptyset,\cF}$ and variable names $\cV$, and let
\begin{align*}
  H=\cV\cup\cC\cup
  \{f(x_1,\ldots,x_r):f\in\cF,\ \rk(f)=r,\ x_1,\ldots,x_r\in\cV\}.
\end{align*}
The shallow sentences considered here use equalities and disequalities between terms in $H$, Boolean constants, $\land$, $\lor$, and existential quantification.
Coherence is imposed on this presentation; no normalization of arbitrary higher-order generated terms is assumed.
For annotations, treat each shallow literal as one leaf.

An \emph{active tree} retains the root, both children of each retained conjunction, the child of each retained quantifier, and exactly one child of each retained disjunction.
All other nodes are ignored.
This makes explicit the branch selection for disjunction in \cite[Appendix~C]{Mathur2025Theories}.
At every active node $u$, choose an equivalence relation $E_u$ on $H$ whose complete equality diagram $\Delta(E_u)$ is satisfiable: this diagram contains $s=t$ when $sE_ut$, and $s\neq t$ otherwise, treating variable names as constants.
There are finitely many such diagrams, and their satisfiability is decidable by finite EUF satisfiability.
The annotation is \emph{locally consistent} if the diagram at each active literal entails that literal, there is no active $\bot$, the diagrams agree across every active Boolean edge, and the following formula is satisfiable at each $u=\exists x.\psi$ with child $v$:
\begin{align*}
  J_{uv}:=\Delta(E_u)\land\Delta(E_v)[x'/x],
\end{align*}
where $x'$ is fresh.
All other variable names and constants remain shared between the two diagrams.
Write $\mathrm{LC}$ for the sentences admitting a locally consistent active annotation.
Every satisfiable sentence belongs to $\mathrm{LC}$, by choosing witnesses, a true disjunct at each active disjunction, and the equality diagrams induced by the resulting assignments.

For a locally consistent annotation, define $\sim$ as the least equivalence relation on pairs $(u,t)$, with $u$ active and $t\in H$, generated by:
\begin{enumerate}
  \item $(u,s)\sim(u,t)$ whenever $sE_ut$;
  \item $(u,t)\sim(v,t)$ across every active Boolean edge;
  \item $(u,s)\sim(v,t)$ across an existential edge as above whenever
  $J_{uv}\vDash s=t[x'/x]$.
\end{enumerate}
A \emph{forgetful path} has endpoints $u,v$ such that some $\sim$-class contains $(u,x)$ and $(v,y)$ for variables $x,y\in\cV$, but contains no $(w,z)$ with $z\in\cV$ at some intermediate node $w$ on that path.
The class still has a shallow-term representative at every node on this path, since the generating edges of $\sim$ connect adjacent nodes of a tree.
Thus a forgetful path loses all variable representatives of a tracked value and subsequently recovers one.
We define $\Coh$ to contain the locally inconsistent shallow sentences and those admitting a locally consistent active annotation without a forgetful path.

\begin{myparagraph}{Why local consistency suffices on $\Coh$}
We first show that $(u,s)\sim(u,t)$ holds exactly when $sE_ut$.
To see the nontrivial direction, take a walk witnessing $\sim$ and successively remove excursions that traverse an edge and return.
At a Boolean edge the two diagrams agree.
At an existential edge, the transport equalities and the child's diagram together entail the resulting equality at the parent; since $J_{uv}$ is satisfiable and contains the parent's complete diagram, that equality must already belong to $E_u$.
The same argument applies to an excursion from child to parent and back.
We thus prove the claim by induction on the walk length.
Using the same argument, we show that $\sim$-equivalent terms at adjacent nodes are related by that edge's transport relation, after applying the local diagrams.

Suppose now that there is no forgetful path and $(u,x_i)\sim(v,y_i)$ for $1\leq i\leq r$.
At every node $w$ on the path from $u$ to $v$, each input class has a variable representative $z_{w,i}$.
Across a Boolean edge, the equality of these representatives follows from the common diagram; across an existential edge, it follows from the joint diagram for that edge, with the appropriate renaming.
Congruence therefore transports the corresponding function applications along the same path, giving
\begin{align*}
  (u,f(x_1,\ldots,x_r))\sim(v,f(y_1,\ldots,y_r)).
\end{align*}
Constants are transported across every edge as well, equivalently treating them as nullary functions.

Take as a universe the $\sim$-classes, adding an arbitrary element if this set is empty.
At node $u$, assign variable $x$ the class of $(u,x)$.
For a tuple represented by variables $x_1,\ldots,x_r$ at a common node $u$, interpret its image under $f$ as the class of $(u,f(x_1,\ldots,x_r))$.
The preceding argument shows that this interpretation is independent of the chosen representatives and node.
Interpret functions arbitrarily on all other tuples and interpret each constant by its common class.
The same-node property makes every active literal true.
Assignments agree across Boolean edges and, across an existential edge binding $x$, on every variable other than $x$.
By induction on the active tree, we verify that this is a model of the original sentence, with the chosen child witnessing each disjunction.
Consequently, a sentence in $\Coh$ is satisfiable if and only if it belongs to $\mathrm{LC}$.
\end{myparagraph}

\begin{myparagraph}{Regularity and transfer}
Local consistency is effectively regular: active choices and diagrams range over finite sets, and all edge and literal tests are computable local tests.
Coherence is effectively regular as well.
Indeed, $\sim$ is reachability in a graph with one copy of each active tree node for each $t\in H$, with edges determined by the finite annotations.
Reachability is MSO-definable by quantifying families of sets closed under these edges.
The path between two tree nodes and the existence of variable representatives along that path are also MSO-definable.
We construct a finite tree automaton for $\Coh$ by existentially quantifying the active choices and diagrams and taking the union with the complement of $\mathrm{LC}$ within the regular language of shallow sentences.
Expanding the finitely many shallow literal labels into their ordinary term trees preserves regularity.
\end{myparagraph}

\coherentSatisfiability*

\begin{proof}
  Construct a finite tree automaton $A_{\mathrm{lc}}$ for $\mathrm{LC}$ over the input signature and variable names.
  For a scheme $G$ generating sentences in $\Coh$, the preceding equivalence shows that its disjunctive reading is satisfiable exactly when
  $\cL(G)\cap\cL(A_{\mathrm{lc}})\neq\emptyset$.
  We decide this using \cref{lem:hors-automaton-nonemptiness} for every fixed $k$, including when $\cL(G)$ is empty.
  If needed, $\cL(G)\subseteq\Coh$ can also be checked using the regular complement of $\Coh$.
\end{proof}

\subsection{Quantifier-free undecidability}\label[appendix]{app:sat-qf-undecidable}

\quantifierFreeUndecidability*

\begin{proof}
  Intersection emptiness for context-free grammars is undecidable~\cite{Forster2018PCP}.
  It remains undecidable over the fixed alphabet $\set{0,1}$: an injective fixed-length binary encoding of any input alphabet preserves both context-freeness and intersection emptiness.
  Let $C_1,C_2$ be two such grammars, with disjoint nonterminal sets and start symbols $A_1,A_2$.
  We may assume that their productions have the forms $A\to BC$, $A\to a$ for $a\in\set{0,1}$, and $A\to\varepsilon$, by the usual elimination of unit productions and binarization.
  Write $L_i=\cL(C_i)$.

  \begin{myparagraph}{Conjunctive reading}
  Use unary function symbols $f_0,f_1$ and a constant $c$, and encode words by $t_\varepsilon=c$ and $t_{aw}=f_a(t_w)$.
  For each grammar nonterminal introduce a scheme nonterminal $A:o\to o$, translating the three production forms respectively as
  \begin{align*}
    A\,u&\to B\,(C\,u),
    &A\,u&\to f_a\,u,
    &A\,u&\to u.
  \end{align*}
  The additional start rule $S\to\neg(A_1\,c=A_2\,c)$ generates exactly
  \begin{align*}
    \cL(G_\land)=\{t_u\neq t_v:u\in L_1,\ v\in L_2\}.
  \end{align*}
  Indeed, the translation composes the word contexts in their original order, and every rule uses its argument exactly once, so reducing a pending argument earlier does not introduce additional words.
  If $w\in L_1\cap L_2$, the generated sentence $t_w\neq t_w$ makes the conjunctive reading unsatisfiable.
  Conversely, if $L_1\cap L_2=\emptyset$, the free term algebra over $c,f_0,f_1$ satisfies every generated disequality.
  Hence $\bigwedge G_\land$ is satisfiable exactly when $L_1\cap L_2=\emptyset$.

  \end{myparagraph}

  \begin{myparagraph}{Disjunctive reading}
  For this construction, use unary function symbols $f,g$ and the five constants $c,e,d_0,d_1,d_\#$.
  The function $f$ advances along a word, $g$ records its letters, and $e$ marks its endpoint.
  Put
  \begin{align*}
    \Delta=(d_0\neq d_1)\land(d_0\neq d_\#)\land(d_1\neq d_\#).
  \end{align*}
  Introduce a nonterminal $A:o\to o\to o$ for each grammar nonterminal.
  Its two arguments denote the endpoints of the word it derives.
  For every production $p:A\to BC$, introduce $P_p:o\to o\to o\to o$ with rules
  \begin{align*}
    A\,u\,v&\to P_p\,u\,v\,u,\\
    P_p\,u\,v\,z&\to P_p\,u\,v\,(f\,z),\\
    P_p\,u\,v\,z&\to (B\,u\,z)\land(C\,z\,v).
  \end{align*}
  The loop chooses a prospective split point $z=f^m(u)$, for some $m\geq0$; the formulae subsequently generated by $B$ and $C$ constrain it to be a valid shared endpoint.
  The terminal and empty productions become, respectively,
  \begin{align*}
    A\,u\,v&\to (g(u)=d_a)\land(f(u)=v),
    &A\,u\,v&\to u=v.
  \end{align*}
  Finally, use the start rule
  \begin{align*}
    S\to\Delta\land(g(e)=d_\#)\land(A_1\,c\,e)\land(A_2\,c\,e),
  \end{align*}
  with any fixed binary bracketing of the conjunctions.
  Denote the resulting scheme by $G_\lor$.

  For $w=a_0\cdots a_{n-1}$ and ground terms $u,v$, abbreviate the finite formula
  \begin{align*}
    E_w(u,v):=(f^n(u)=v)\land
      \bigwedge_{0\leq j<n}(g(f^j(u))=d_{a_j}).
  \end{align*}
  Every finite terminal derivation from $A\,u\,v$ projects to a grammar derivation $A\Rightarrow^*w$: a completed $P_p$ loop selects the binary production $p$ and its two subderivations.
  We claim that the generated formula entails $E_w(u,v)$ in every structure.
  The terminal and empty cases are immediate.
  At a binary production, we obtain $E_r(u,z)$ and $E_s(z,v)$ for the two derived words $r,s$ by the inductive hypotheses.
  Substituting the equality $f^{\size{r}}(u)=z$ into the latter formula, we obtain $E_{rs}(u,v)$.
  This argument requires neither that the chosen exponent $m$ equal $\size{r}$ nor that $f$ be injective: the generated equalities ensure that the split point denotes the endpoint of the word $r$.

  Suppose that a model satisfies a sentence generated by $G_\lor$.
  Using the preceding invariant, we obtain words $u\in L_1$, $v\in L_2$ such that the model satisfies $E_u(c,e)$ and $E_v(c,e)$, as well as $\Delta$ and $g(e)=d_\#$.
  If the words disagree at some position $j$ before either ends, then $g(f^j(c))$ equals both $d_0$ and $d_1$, contradicting $\Delta$.
  Otherwise, if $u$ is a proper prefix of $v$, then $f^{\size{u}}(c)=e$ makes $g(f^{\size{u}}(c))$ equal to $d_\#$, while the next letter of $v$ makes it equal to $d_0$ or $d_1$, again contradicting $\Delta$.
  The symmetric case is identical, so $u=v$ and $L_1\cap L_2\neq\emptyset$.

  Conversely, let $w=a_0\cdots a_{n-1}\in L_1\cap L_2$ and fix a derivation tree in each grammar.
  At each binary production whose left subtree derives $r$, choose exactly $\size{r}$ iterations of its $P_p$ loop.
  This produces two finite formulae whose split terms describe the corresponding positions in $w$.
  To satisfy their conjunction, take the domain
  \begin{align*}
    \{0,\ldots,n\}\mathbin{\dot\cup}\{\ell_0,\ell_1,\ell_\#\},
  \end{align*}
  interpret $c=0$, $e=n$, and $d_b=\ell_b$ for $b\in\{0,1,\#\}$, and set
  \begin{align*}
    f(i)&=i+1 &&(0\leq i<n),
    &f(n)&=n,\\
    g(i)&=\ell_{a_i} &&(0\leq i<n),
    &g(n)&=\ell_\#.
  \end{align*}
  Extend both functions arbitrarily on the three marker elements.
  By induction on either chosen parse tree, we show that each subtree spanning positions $i$ through $j$ generates a true formula when its arguments denote $i,j$.
  Thus the root sentence is true, proving that $\bigvee G_\lor$ is satisfiable exactly when $L_1\cap L_2\neq\emptyset$.

  The construction includes $w=\varepsilon$, where $c$ and $e$ may coincide.
  If either input language is empty, neither scheme generates any sentence; the conjunctive reading is then true and the disjunctive reading false, as required.
  In the disjunctive construction every substituted argument is a terminal ground term, since the start arguments are constants and updates only apply $f$.
  Duplicating these arguments therefore never duplicates an unresolved nondeterministic choice.
  Both schemes have order at most $1$, use fixed finite signatures with two unary functions, and generate closed quantifier-free sentences without relation symbols other than equality.
  The reductions consequently apply at every order $k\geq1$.
  \end{myparagraph}

\end{proof}

\subsection{Regular atomic components}\label[appendix]{app:sat-regular-atoms}

\regularAtomicComponents*

\begin{proof}
  Let $(G,(A_i)_{i\in I})$ be the input presentation, where $G$ generates Boolean skeletons with interface symbols $a_i$, and each $A_i$ is a nonempty regular language of ground atomic formulae.
  We adapt the finite classification of atomic components in the proof of \cite[Theorem~1.5]{Mathur2025Theories}; only the test on Boolean skeletons changes.

  A \emph{profile} assigns each $i\in I$ a nonempty set $p_i\subseteq\set{0,1}$ of permitted truth values.
  There are $3^{\size{I}}$ profiles.
  For a profile $p$, form the ground quantifier-free theory
  \begin{align*}
    U_p
      = \bigcup_{i:p_i=\{1\}} A_i
        \;\cup\!
        \bigcup_{i:p_i=\{0\}}\{\neg\alpha:\alpha\in A_i\}.
  \end{align*}
  This language is effectively regular, by closure under finite union and adjoining a negation at the root.
  Its conjunctive satisfiability is decidable by \cite[Theorem~1.5]{Mathur2025Theories}, including arbitrary finite signatures with relations, functions, and equality.
  Components with $p_i=\set{0,1}$ impose no constraint.

  Next construct a deterministic finite tree automaton that computes, for each Boolean skeleton $t$, the set $V_p(t)$ of its possible truth values when each occurrence of $a_i$ independently receives a value in $p_i$.
  Its states are the three nonempty subsets of $\set{0,1}$, and its transitions are
  \begin{align*}
    V_p(a_i)&=p_i,
    &V_p(\bot)&=\{0\},
    &V_p(\top)&=\{1\},\\
    V_p(\neg t)&=\{1-b:b\in V_p(t)\},\\
    V_p(t\mathbin\circ u)
      &=\{b\mathbin\circ c:b\in V_p(t),\ c\in V_p(u)\}
      &&\text{for }\circ\in\{\land,\lor\}.
  \end{align*}
  Here Boolean operations on $0$ and $1$ have their usual meanings.
  We establish the claimed interpretation of $V_p(t)$ by induction on $t$, using independent choices in the two subtrees at each binary connective.
  We designate the states containing $0$ as accepting to obtain an automaton $B_p$ for the skeletons that can evaluate to false.
  By \cref{lem:hors-automaton-nonemptiness}, we can decide whether
  \begin{align*}
    \cL(G)\cap\cL(B_p)=\emptyset.
  \end{align*}
  This test uses the original order-$k$ scheme; no increase in order is needed.
  The algorithm accepts precisely when some profile passes both this test and the conjunctive satisfiability test for $U_p$.

  For completeness, suppose that a model $\cM$ satisfies every generated formula.
  Choose the exact profile
  \begin{align*}
    p_i=\{\mathbf 1_{\cM\vDash\alpha}:\alpha\in A_i\}.
  \end{align*}
  Each $p_i$ is nonempty, and $\cM\vDash\bigwedge U_p$.
  Fix any $t\in\cL(G)$ and any permitted assignment of truth values to its interface occurrences.
  At each occurrence of $a_i$, choose an atom from $A_i$ having the assigned truth value in $\cM$.
  Independent substitution makes the resulting formula an element of the generated theory, so it is true in $\cM$.
  Its truth value is exactly that of the chosen Boolean evaluation of $t$.
  Thus $V_p(t)=\{1\}$ for every generated skeleton, and the second test also succeeds.

  For soundness, let $p$ pass both tests and choose a model $\cM\vDash\bigwedge U_p$.
  Every atom in $A_i$ has a truth value belonging to $p_i$: a singleton $p_i$ forces the specified value, while a two-element $p_i$ permits either value.
  Consequently, any formula obtained by substituting atoms into $t\in\cL(G)$ has its truth value in $V_p(t)$.
  The second test ensures that $V_p(t)=\{1\}$, so $\cM$ satisfies every generated formula.
  Since the finite search and both tests terminate, we conclude decidability.
\end{proof}

\end{document}